# Homogeneous Nucleation of Phase transformations in Supercooled Water

Robert F. Tournier

*Univ. Grenoble Alpes, CNRS, Grenoble INP\*, Institut Néel. 38000 Grenoble, France*
*\*Institute of Engineering Univ. Grenoble Alpes*

e-mail: robert.tournier@neel.cnrs.fr

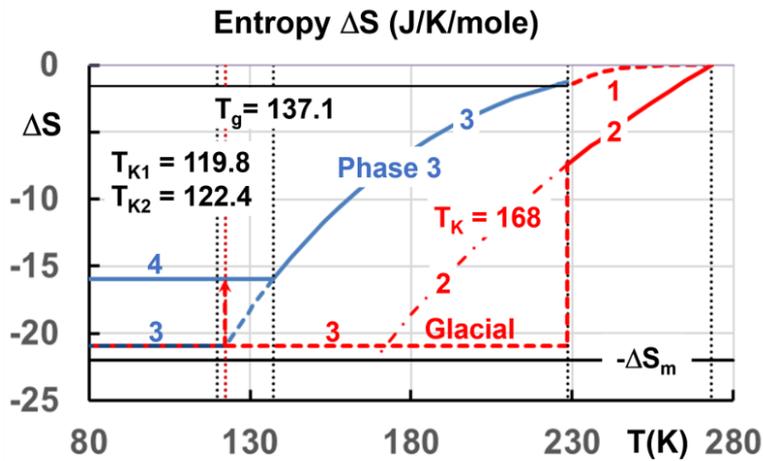

Entropy ΔS (J/K/mole) of water and glacial Phases:
1- Disordered Phase 3 entropy
2- Water entropy and its Kauzmann temperature $T_K \cong 168$
3- 1st cooling with 1st-order transition, leading to the lowest entropy $\Delta S = -20.967$ at $T_{LL}$ and first-order transition at $T_{K2} = 122.4$ without latent heat
3- 2nd cooling: Phase 3 formation and underlying 1st order transition $T_{K2} = 122.4$ K; $T_K = 119.8$ K
4- Glass Phase 3 with $T_g = 137.1$

**Abstract**: The classical nucleation equation, applied to two liquids, is completed by an additional enthalpy for solid supercluster formation governing the liquid and glass transformations. This model defines a formation rule of glacial phases, explaining the origin of the first-order transition of water from fragile-to-strong liquids at $T_{LL} = 228.5$ K, only fixing $T_g = 137.1$ K. All thermodynamic properties and transitions, even under pressure P, are now predicted in agreement with experiments. The lowest-density liquid is, at once, the glacial phase of fragile and high-density liquids. It is formed at $T_{LL}$, remains liquid during the first cooling, and gives rise to the glass Phase 3 by heating through a first-order transition without latent heat at $T_{K2} = 122.4$ K. Medium-range order appears above $T_g$ during the first heating. Phase 3 with its own Kauzmann temperature undergoes the first-order transition at $T_{LL}$ without latent heat and superheats up to $T_{n+} > T_m$.

**Keywords**:

1_ **undercooled liquids and glasses**;

2_ **homogeneous nucleation**;

3_ **glacial phases**;

4_ **"Organized" Liquid**;

5_ **First-order transitions.**

**Acronyms**:

HDA: High-density amorphous

LDA: Low-density amorphous

HDG: High-density glass

LDG: Low-density glass

VHDA: Very high-density amorphous

LLPT: Liquid-liquid phase transition

VFT: Vogel-Fulcher-Tammann

Chap.: Chapter

Eq.: Equation

Eqs.: Equations

**Introduction**

Amorphous ice is obtained by vapor condensation at low temperatures, rapid cooling of emulsion, confined water or applying high pressure. Water can be supercooled down to 235 K before entering the first ice crystallization zone. Although liquid water does not vitrify upon continuous cooling, an amorphous form of ice was first obtained in 1935 by vapor deposition on a copper surface kept below 163 K [1]. In 1965, an experimental device for depositing vitreous ice and performing differential thermal analysis in situ detected a glass transformation at 139 K followed by spontaneous crystallization in a second zone [2]. In 1968, a high heat capacity jump was measured at 133.5 K by heating a vapor-deposited bulk sample of 0.6 g followed by crystallization [3]. Heat capacity measurements of thin films lead to much smaller jumps around 130 K and very often to glass transition temperatures of 179-180 K [4-7]. The variations of the glass transition temperature with composition for various binary solutions of which water is the common component, have confirmed that $T_g$ is 135–140 K [8]. Water was still supercooled when it was confined within silica-gel voids with an average diameter of 1.1 nm [9]. Two apparent glass transitions were observed around 135 and 173 K. The two crystallization zones separated by ≈100 K were inevitably produced in two different liquids separated by a transformation occurring in this range of temperatures. A liquid-liquid transition was observed at around $T_{LL}$ = 227-228 K showing a heat capacity bump of 98-110 J/K/mol of confined water and emulsion [9]. The water heat capacity strongly increased in the temperature range from 300 to 235 K [9-13].

In 1984, the ice-to-water transformation associated with the decrease under pressure of the melting temperature $T_m$ was used to supercool bulk water below $T_g$ [14]. Two coexisting liquids of different densities HDA and LDA were separated at various temperatures thanks to first-order transitions under pressures lower than ≈ 0.6 GPa [15-18]. A second transition was produced under ≈ 0.95 GPa leading to a higher density phase called VHDA [19].

Many theoretical and computational studies attribute the two-liquid separation to the existence of a first-order transition line $T_{LL}(P)$ under pressure P that terminates at a critical point [20-27]. This transition line could be due to a separation of two supramolecular arrangements of hydrogen bonds in water. The main difficulty is to determine it within the no man's land of the first crystallization zone.

There are several descriptions of glass transition as a true phase transformation. A typical model is based on a percolation-type phase transition with the formation of dynamic fractal structures close to the percolation threshold [28,29,30]. Macroscopic percolating clusters formed at the glass transition have been visualized [31]. High precision measurements of fifth-order, non-linear dielectric susceptibilities of growing transient domains diverge towards $T_g$ [32]. Excited delocalized atoms would be responsible for viscous flow and time dependence of $T_g$. At the glass transition, the process of atom delocalization would be reduced without being fully eliminated [33]. Theoretical works based on the notion that a random first-order transition lies at the heart of glass formation have been developed. The possible existence of an underlying first-order transition without latent heat has been raised [34-37]. Its presence in many glass-forming melts has been shown [38].

The glass phase is viewed as having a thermodynamic origin and resulting from the initial existence of two liquid phases in water [20] and in several liquids [39]. The classical nucleation equation is completed for each liquid by an additional enthalpy, predicting the existence of several homogeneous nucleation temperatures [40,41]. Liquid 1 and Liquid 2 have different homogeneous nucleation temperatures of superclusters: $T_1$, within the no man's land, which is not easily observed and $T_2$, a lower temperature where superclusters percolate and interpenetrate. The transition at ($T_2$) is accompanied by an enthalpy change from Liquid 1 to Liquid 2 measured by heating associated with the formation of the glass phase at $T_g = T_2$ [40]. An underlying first-order transition temperature without latent heat, gives rise, after the first cooling, to a new phase that I call Phase 3 with an enthalpy equal to the difference between those of Liquids 1 and 2. This phase is 'ordered' at the temperature $T_{K2}$ in all glasses where the enthalpy difference between Liquids 1 and 2 is exactly opposite to that predicted at $T_m$. This new supercooled phase is a direct consequence of the presence of two liquids and is an "ordered" liquid above $T_g$ which superheats up to $T_{n+}$ where superclusters melt through a first-order transition. These superclusters and the "ordered" Phase 3 revive during a new cooling at another homogeneous nucleation temperature $T_{n+}$ below $T_m$ [38].

A new glass phase called glacial phase has been induced in 1995 in triphenyl phosphite by an isotherm annealing of limited duration above $T_g$ [42,43]. The origin of this phase is attributed to the presence of two liquids [38,44]. Other glacial phases have been discovered in n-butanol and d-mannitol [45,46]. They undergo a liquid-to-glass first-order transition at a temperature much higher than their initial $T_g$, leading to a new glass phase of lower enthalpy when rapid heating and cooling are applied to prevent crystallization. Its enthalpy cannot be lower than a minimum enthalpy defined by the initial Phase 3. The melting temperature of this new glass phase occurs when its entropy change can be accommodated by the entropy of the new liquid Phase 3 existing above it [38] and is observed at high heating rates.

This model also explains the presence of intrinsic nuclei above $T_m$ submitted to the Laplace pressure of liquid [47,48], predicts the latent heat of the first-order transition from liquid helium under pressure to a glass phase [49] and the Lindemann constant of pure liquid elements [50].

This model has been applied to water [51] and the transition at $T_{LL}$ is viewed as a transformation of fragile-to-strong liquid, as suggested in 1999 [52]. Glass-to-glass and liquid-to-liquid transformations of water under pressure and their latent heats are predicted. A supercooled liquid called Phase 3 is formed below $T_g$ thanks to the difference of additional enthalpies of Liquids 1 and 2. The transition at $T_{LL}$ during heating does not appear as a first-order transition because liquid Phase 3 and Liquid 1 are submitted to enthalpy changes of opposite sign [51,53].

In this new publication, the glass transition temperature is $T_g = 137.12$ K, and the reduced first-order transition temperature is now fixed at $\theta_{LL} = (T_{LL}-T_m)/T_m = -0.1633$. This temperature $T_{LL}$ is viewed as the formation temperature in two fragile liquids of a glacial and strong Phase 3 by cooling. This view has as for consequence that the latent heat does not vary with pressure at this temperature and a new phase diagram P(T) is revealed. In addition, an underlying first-order transition occurs at $T_{K2} = 122.4$ K without latent heat as theoretically suggested [34-36], above the Kauzmann temperature $T_{K1} = 119.8$ K transforming Phase 3 into an "ordered" glacial phase by an enthalpy decrease. The specific heat, the transition temperatures, the entropy and enthalpy of various phases are calculated, for the first time, as a function of the temperature and pressure in agreement with many experimental results [1,3,6,7,15,16,19,54-59] by only knowing the glass transition $T_g$, the melting temperature $T_m$, and the melting enthalpy $\Delta H_m$.

In this paper, the word "glass" is used for a thermodynamic phase having a well-defined and constant enthalpy and entropy below $T_g$ after a heat capacity jump. The word "amorphous" is used by experimentalists, observing low and high-density water transformations. The glass phase appears below the glass transition temperature $T_g$ of the amorphous state.

## 2. Predicting glass-to-glass and liquid-to-liquid transitions

### 2.1. *Nucleation temperatures $T_1$ and $T_2$ of "ordered" Liquids 1 and 2 and the glass transition at $T_g = T_2$*

Many experimental results show that nuclei survive above the melting temperature due to the existence of Laplace pressure varying as the radius inverse $R^{-1}$ [39,40,47,48,60,61]. The starting point of the present thermodynamic analysis of amorphous ices in terms of two initial liquid phases is the classical nucleation theory as complemented for each liquid by an additional enthalpy term that gives rise to nuclei surviving in water above $T_m$. This complementary enthalpy $\varepsilon \Delta H_m/V_m$ is added to the classical Gibbs free energy change $\Delta G$ associated with the formation of a nucleus of radius R in Eq. (1):

$$\Delta G = \frac{4\pi R^3}{3\, V_m} \Delta H_m \times (\theta - \varepsilon) + 4\pi R^2 (1+\varepsilon)\sigma_1, \qquad (1)$$

where ($\varepsilon$) is a fraction of the melting enthalpy $\Delta H_m$ per mole (equal to $\varepsilon_{ls}$ for a nucleus in Liquid 1, $\varepsilon_{gs}$ for a nucleus in Liquid 2, $\Delta\varepsilon_{lg} = (\varepsilon_{ls} - \varepsilon_{gs})$ for a nucleus of the supercooled Phase 3), $V_m$ is the molar volume, $\sigma_1$ the surface energy in the classical nucleation equation and $\theta = (T-T_m)/T_m$ is the reduced temperature. The melting heat $\Delta H_m$ and $T_m$ are assumed to be the same for all phases and do not depend on R, regardless of the nucleus radius. The nuclei are superclusters instead of tiny crystals because the transformation temperatures are observed in liquids and give rise to classical or ultrastable glass phases or even to new liquid phases.

At thermodynamic equilibrium, the derivative $d(\Delta G)/dT$ at $T_m$ of Eq. (1) is always equal to the melting entropy $\Delta S_m$. The volume contribution to Eq. (1) in the classical nucleation equation is equal to $(-n\Delta\mu)$, where n is the atom or molecules number, $\Delta\mu$ the difference of chemical potentials $(\mu_l - \mu_s)$ per atom in the liquid and the supercluster phase that is written in Eq. (3) as a Taylor expansion of $\Delta\mu(T)$ in the vicinity of the melting temperature [62] including Eq. (2) in $\Delta\mu$:

$$\varepsilon = \varepsilon_{\dot{o}}\left(1 - \frac{\theta^2}{\theta_{0g}^2}\right), \quad (2)$$

$$\Delta\mu(T) = \frac{\Delta H_m}{n}\left[1 + \varepsilon_0 - \frac{\varepsilon_0}{\theta_0^2} - \frac{T}{T_m}\left(1 - 2\frac{\varepsilon_0}{\theta_0^2}\right) - \left(\frac{T}{T_m}\right)^2 \frac{\varepsilon_0}{\theta_0^2}\right]. \quad (3)$$

$\Delta\mu$ is equal to $\varepsilon_0\Delta H_m/n$ at $T/T_m = 1$ and equal to zero at a new temperature respecting $\theta = \varepsilon$ instead of $\theta = 0$. In these conditions, a supercluster phase survives above $T_m$, and melts at $\theta = \varepsilon$ accompanied by an endothermic heat equal to $\varepsilon \times \Delta H_m$ if crystallization is avoided. The existence in the liquid of "an isolated avoided critical point" at a temperature $T^* > T_m$ has already been considered in the past in one-liquid model leading to a locally preferred structure which differs greatly from the local structure in the actual crystalline phase [63].

The existence of $\varepsilon_0$ suggests the presence of an underlying first-order transition, without latent heat, in the supercooled liquid, limiting its enthalpy decrease at low temperatures to $(-\varepsilon_0\Delta H_m)$ per mole in the absence of glass transition [34-38]. The value $\varepsilon_0$ is known for pure liquid elements and equal to 0.217 [39] and is successfully used to determine the Lindemann's constant 0.103 [50]. Temperatures where $\theta = \varepsilon$ have been observed [38,60,64,65].

The glass nuclei are spontaneously formed by homogeneous nucleation in the supercooled liquid and melted above $T_m$ by homogeneous nucleation instead of surface melting. They are known today as solid superclusters very often having an icosahedral structure [66-69]. The critical nuclei give rise to "ordered" Liquids 1 and 2 below their own homogeneous nucleation temperature $T_1$ and $T_2$. The two ordered liquids are transformed into glass Phase 3 below $T_g = T_2$, or various LLPT above $T_g$ and even $T_m$, according to the thermal variations of $\varepsilon$ [41]. The new surface energy is $(1+\varepsilon)\sigma_1$ instead of $\sigma_1$ because the classical equation is obtained for $\varepsilon = 0$ [70]. The homogeneous nucleation temperatures $\theta_{n-}$ and $\theta_{n+}$ of these phases are given in Eqs. (4,5) and the thermally activated critical energy barrier in Eq. (6) [6,39,40]:

$$\theta_{n-} = \frac{\varepsilon - 2}{3}, \quad (4)$$

$$\theta_{n+} = \varepsilon, \quad (5)$$

$$\Delta G^* / (k_B T) = \frac{12(1+\varepsilon)^3 Ln(K.v.t) / \frac{81}{(\theta - \varepsilon)^2}}{(1+\theta)}, \quad (6)$$

where v is the sample volume and t the nucleation time, Ln (K.v.t) $\cong$ 90.

The critical radius and the thermally activated energy barrier are infinite at the homogeneous nucleation temperatures obtained for $\theta_{n+} = \varepsilon$ instead of $\theta = 0$ for the classical equation. A catastrophe of nucleation is predicted at the superheating temperature $\theta_{n+} = \varepsilon$ for crystals protected against surface melting or for the "melting" of various "ordered" liquids during fast overheating [38,71]. Each liquid is "ordered" at $\theta = \theta_{n-} < 0$ and this order disappears by superheating at $\theta = \theta_{n+}$.

The coefficients $\varepsilon_{ls}$ and $\varepsilon_{gs}$ in Eqs. (7-8) represent values of $\varepsilon(\theta)$, and lead to the nucleus formation with a critical radius for "ordered" Liquid 1 and Liquid 2 under pressure:

$$\varepsilon_{ls}(\theta) = \varepsilon_{ls0}\left(1 - \theta^2/\theta_{0m}^2\right) + \Pi_1, \tag{7}$$

$$\varepsilon_{gs}(\theta) = \varepsilon_{gs0}\left(1 - \theta^2/\theta_{0g}^2\right) + \Pi_2 + \Delta\varepsilon, \tag{8}$$

where $\Pi_1 = (P-P_0)V_{m1}/\Delta H_m$ and $\Pi_2 = (P-P_0)V_{m2}/\Delta H_m$ are the contributions of the pressure (P) to the enthalpy coefficients $\varepsilon_{ls}$ and $\varepsilon_{gs}$, $P_0$ the ambient normal pressure equal to 0.0001 GPa, $\Delta\varepsilon$ an enthalpy excess coefficient in Liquid 2 due to the latent heat of a first-order transition toward a lower enthalpy phase. The dependence of $\varepsilon$ on $\theta^2$ is deduced from the maximum supercooling rate of 30 liquid elements [39]. The free volumes in Liquid 1 and Liquid 2 are equal to zero when the coefficients $\varepsilon_{ls}$ and $\varepsilon_{gs}$ are equal to zero at the reduced temperatures $\theta_{0m}$ and $\theta_{0g}$ for $\Pi_1 = 0$ and $\Pi_2 = 0$ which roughly correspond to the VFT temperatures of "ordered" Liquid 1 and Liquid 2 respectively, even if viscosity descriptions show that the infinite viscosity does not occur in real systems [72-75]. The free volume is well defined when the superclusters are completely formed below the homogeneous nucleation temperature for $\theta_g < \theta < \theta_1$ and is progressively reduced by decreasing the temperature. The VFT temperature could be equal to $T_{0m}$ during the first cooling in this window of temperatures. The infinite value of viscosity cannot be attained at $T_{0m}$ because the homogeneous nucleation of superclusters in Liquid 2 occurs at $\theta_g$ and tends to fill the free volume. The enthalpy of "ordered" Liquid 1, in the absence of glass transition, would be equal to that of the initial liquid without superclusters at the temperature $T_{0m}$. There is no more free volume at $T_{0m}$ in this amorphous solid. It is impossible to delocalize any atom at this temperature [33].

Eqs. (7-8) are applicable at the homogeneous nucleation temperatures $\theta_{n-}$ given by Eq. (4) for ordering formation in Liquids 1 and 2. Equation (9) determines $\theta_{n-}$ for Phase 2, combining Eqs. (4,8):

$$\varepsilon_{gs0}\theta_{0g}^{-2}\theta_{n-}^{2} + 3\theta_{n-} + 2 - \varepsilon_{gs0} - \Pi_2 - \Delta\varepsilon = 0. \tag{9}$$

The solutions for $\theta_{n-}$ are given by Eq. (10):

$$\theta_{n-} = \frac{\left(-3 \pm \left[9 - \frac{4(2-\varepsilon_{gs0}-\Pi_2-\Delta\varepsilon)\varepsilon_{gs0}}{\theta_{0g}^2}\right]^{\frac{1}{2}}\right)\theta_{0g}^2}{2\varepsilon_{gs0}}, \tag{10}$$

where $\theta_{n-}$ of "ordered" Liquid 2 for the sign + is $\theta_2 = \theta_g$ and $\theta_{n-}$ of "ordered" Liquid 2 for the sign (−) corresponds to the departure temperature of the enthalpy relaxation and to the glass formation after liquid hyperquenching [38].

Equation (11) determines the homogeneous nucleation temperature $\theta_{n-}$ in Liquid 1 combining Eq. (7) at this temperature with Eq. (4):

$$\theta_{n-}^2 \times \varepsilon_{ls0} \times \theta_{0m}^{-2} + 3\theta_{n-} + 2 - \varepsilon_{ls0} - \Pi_1 = 0. \tag{11}$$

The reduced homogeneous nucleation temperatures $\theta_{n-}$ of Liquid 1 under pressure in Eq. (12) are solutions of Eq. (11):

$$\theta_{n-} = \frac{\left(-3 \pm \left[9 - \frac{4(2-\varepsilon_{ls0}-\Pi_1)\varepsilon_{ls0}}{\theta_{0m}^2}\right]^{\frac{1}{2}}\right)\theta_{0m}^2}{2\varepsilon_{ls0}}, \tag{12}$$

where $\theta_{n-}$ is called $\theta_1$ for the sign +. The first transition leading to "ordered" Liquid 1 occurs by cooling at $\theta_1$ in the no man's land below which Liquid 1 contains superclusters moving in the free volume without percolation. The most important transition occurs at room pressure at the homogeneous nucleation temperature $\theta_g = \theta_2$ of Liquid 2 where new superclusters of Liquid 2 and those of Liquid 1 percolate. This transition observed by heating is accompanied at $\theta_g$ by an enthalpy change from "ordered" Liquid 1 to glass governed by the difference in enthalpy $\Delta\varepsilon_{lg}$ = ($\varepsilon_{ls}$ -$\varepsilon_{gs}$). The boundary $T_{0m} = T_m/3$ separates strong and fragile liquids completing the distinction established by Angell in 1995 [76]. Any strong Liquid 1 has a temperature $T_{0m}$ ( ≈ $T_{VFT}$) lower than or equal to $T_m/3$ while that of any fragile Liquid 1 is larger than $T_m/3$. The two families have different thermodynamic properties below $T_g$. A liquid is fragile when Eqs. (9,11) have a unique solution for $\Pi_1$ and $\Pi_2 = 0$.

## 2.2. Determining coefficients $\varepsilon_{ls0}$, $\varepsilon_{gs0}$, $\theta_{0m}$ and $\theta_{0g}$

In all fragile and strong liquids, the coefficients ($\varepsilon_{gs0}$) in Eq. (13) and ($\varepsilon_{ls0}$) in Eq. (14) deduced from Eqs. (9,11) are calculated with $\Pi_2 = 0$, $\Pi_1 = 0$ as well as the value of $\theta_g$ in the absence of pressure and enthalpy excess coefficient $\Delta\varepsilon$:

$$\varepsilon_{gs0} = \frac{3\theta_g + 2 - \Delta\varepsilon - \Pi_2}{1 - \frac{\theta_g^2}{\theta_{0g}^2}}, \tag{13}$$

$$\varepsilon_{ls0} = \frac{3\theta_g + 2 - \Pi_1}{1 - \frac{\theta_g^2}{\theta_{0m}^2}}. \tag{14}$$

The new glass transition temperature $\theta_g$ depends on $\Delta\varepsilon$ and $\Pi_2$ and is determined from the knowledge of $\theta_2$, $\theta_{0g}$, and $\theta_{0m}$ [40]. The glass transition occurs in strong liquids at the temperature $\theta_g = \theta_{Br}$ respecting $\varepsilon_{ls} = \varepsilon_{gs}$. In the great majority of strong liquids, ($\theta_{0g}$) is equal to (-1) because the relaxation time follows an Arrhenius law below $T_2 = T_g$.

In fragile liquids, the enthalpy difference coefficient $\varepsilon_{ls0}$ is maximized in Eq. (15) for $\Pi_1 = 0$ and $\theta_1 > \theta_g$:

$$\varepsilon_{ls}(\theta = 0) = \varepsilon_{ls0} + \Pi_1 = 1.5 \times \theta_1 + 2 = a \times \theta_g + 2 + \Pi_1, \tag{15}$$

where a = 1 leads to a well-known specific heat excess ($\Delta C_p(T)$) of the supercooled melt at the glass transition equal to $1.5\Delta H_m/T_m$ [40,77]. The temperature $\theta_1$ is equal to $a \times \theta_g/1.5$ for $\Pi_1 = 0$. The reduced temperature ($\theta_{0m}$) is given by Eq. (16), leading to a unique solution for Eq. (8):

$$\theta_{0m}^2 = \frac{8}{9}\varepsilon_{ls0} - \frac{4}{9}\varepsilon_{ls0}^2. \tag{16}$$

New parameters ($\varepsilon_{gs0}$) and ($\theta_{0g}$) are fixed in Eqs. (17,18) to lead to a unique solution for Eq. (9). The enthalpy difference coefficient ($\varepsilon_{gs0}$) is maximized by Eqs. (17,18) for $\Pi_2 = 0$:

$$\varepsilon_{gs}(\theta = 0) = \varepsilon_{gs0} + \Pi_2 = 1.5 \times \theta_g + 2 + \Pi_2 + \Delta\varepsilon, \tag{17}$$

$$\theta_{0g}^2 = \frac{8}{9}\varepsilon_{gs0} - \frac{4}{9}\varepsilon_{gs0}^2. \tag{18}$$

The glass transition of fragile liquids occurs at a temperature $\theta_2 = \theta_g$, lower than that for which $\epsilon_{ls} = \epsilon_{gs}$. A second glass transition is expected for $\Delta\epsilon_{lg} = 0$ when an enthalpy excess coefficient $\Pi$ is added under pressure to those of Liquids 1 and 2 and induces a first-order glass transition defined by Eq. (10) with a latent heat coefficient $\Pi_2$ leading to this second value of $T_g$ respecting $\Delta\epsilon_{lg} = 0$ with $\Delta\epsilon = 0$.

## 2.3. *Formation of* a *liquid ordering at homogeneous nucleation temperatures, supercooling and superheating of a new phase called Phase 3*

Elementary superclusters containing a critical number n of atoms given in Eq. (19) are formed in liquids 1 and 2 at the homogeneous nucleation temperatures $T_1$ and $T_2$ during the first cooling [40]:

$$n = -8N_A k_B (1 + \varepsilon)^3 LnK / [27\Delta S_m (\theta - \varepsilon)^3] \tag{19}$$

The value of $(1+\epsilon_{gs})/(\theta_2-\epsilon_{gs})$ in Eq. (19) is 1.5 and n = $751/\Delta S_m \cong 102$ (or about 34 water molecules), assuming $LnK \cong 90$ and $\Delta S_m = 7.322$ J/K/g-atom, the melting entropy per g-atom [53].

The homogeneous nucleation temperatures $T_1$ and $T_2$ are defined for liquids which are free of nuclei before supercooling. Superclusters in Liquid 1 move in the free volume and give rise to an "ordered" Liquid 1 characterized by short-range order below $T_1$ without volume and enthalpy changes at $T_1$. At the homogeneous nucleation temperature $T_2 = T_g$ in Liquid 2, all superclusters in liquids 1 and 2 percolate. This event does not lead to freezing below $T_g$ because there is no enthalpy difference between those of liquids 1 and 2 during the first cooling as shown by hyperquenching [38]. Consequently, Liquid 2 has an enthalpy excess equal to $\Delta\epsilon = (\epsilon_{ls0} - \epsilon_{gs0})$ at temperatures lower than the temperature $T_{K2}$ where a first-order transition without latent is expected. Glass phase begins to be built during heating at the second homogeneous nucleation temperature predicted in Eq. (10) with the sign (-) for $\Delta\epsilon = \Delta\epsilon_{lg0} = (\epsilon_{ls0} - \epsilon_{gs0})$, $\Pi_2 = 0$ in fragile liquids and at the temperature $T_{K2}$ in strong liquids. The enthalpy excess is fully recovered during heating and a specific heat jump is now observed at $T_g$. The glass phase is then broken in multiple blocks at $T_g$ during heating when the broken bonds called configurons percolate and give rise to a new "ordered" liquid phase with medium-range order made of blocks of atoms moving in the free volume, and containing many elementary superclusters [28].

The "melting" of any "ordered" liquid phase made of atom blocks or of elementary superclusters moving in the free volume occurs at $\theta_{n+} > 0$ defined in Eq. (5) and the "ordering" could be recovered at $\theta_{n+} < 0$. This order disappears after overheating at $T_{n+}$ above $T_m$ [38] and can reappear by slow cooling below $T_m$ at a second temperature $T_{n+}$ which is symmetrical from the first one regarding $T_m$. This phenomenon has been observed for $Zr_{41.2}Ti_{13.8}Cu_{12.5}Ni_{10}Be_{22.5}$ using electrostatic levitation and for tin droplets by using a very high cooling rate [64,65]. Melting of residual nuclei has been observed after an overheating of 247 K above the melting temperature of eutectics CoB [60]. The order reappears because the applied overheating rate above $\theta_{n+} > 0$ is too weak to melt all residual superclusters of very small size [48]. Their melting at $\theta_{n+} > 0$ is uncomplete due to the higher density of their core [78]. Consequently, after melting, each supercluster could be replaced by a supercluster containing a much smaller magic atom number acting as critical nucleus at $\theta_{n+} < 0$.

The enthalpy difference coefficient ($\Delta\varepsilon_{lg}$) between Liquids 1 and 2 given by Eq. (20) under pressure gives rise to the new glass phase below $\theta_g$ that I called Phase 3 instead of "ordered" Liquid 2 [51]:

$$\Delta\varepsilon_{lg(\theta)} = \varepsilon_{ls} - \varepsilon_{gs} = \varepsilon_{ls0} - \varepsilon_{gs0} + \Delta\varepsilon + \Pi_1 - \Pi_2 - \theta^2 \left( \varepsilon_{ls0}/\theta_{0m}^2 - \varepsilon_{gs0}/\theta_{0g}^2 \right), \quad (20)$$

where $\Delta\varepsilon$ is here a latent heat coefficient associated with a first-order transition (above $T_g$ for glacial phase and below $T_g$ for ultrastable phase) or corresponds to an enthalpy excess obtained by hyperquenching or vapor deposition below $T_g$. Phase transformations are produced when $\Delta\varepsilon_{lg}$ becomes equal to zero. The thermally–activated energy barrier, defined by Eq. (21) for the transformation of Phase 3 [40], becoming infinite at $\theta = \theta_{sg}$ for $\Delta\varepsilon_{lg} = 0$, can give rise to a first-order transition:

$$\Delta G^*/(k_B T) = 12(1 + \Delta\varepsilon_{lg})^3 \frac{Ln(K.v.t)}{81(1+\theta)\Delta\varepsilon_{lg}^2} \quad . \quad (21)$$

The excess $\Delta\varepsilon$ obtained after hypercooling has a maximum value $\Delta\varepsilon_{lg0}$, which increases $\Delta\varepsilon_{lg}$ up to zero at the underlying first-order transition temperature $T_{K2}$ [38]. The difference $\Delta\Pi_1 = \Pi_1-\Pi_2 = \delta V P/\Delta H_m$ acting in Eq. (20) is also positive because $dT_m/dP$ is negative and increases with the volume change $\delta V$ and the pressure P up to the transformation temperature $T_{sg}$, where $\Delta\varepsilon_{lg}(\theta)$ still attains zero for $\Delta\varepsilon = 0$. The pressure contribution ($\Delta\Pi$) is still equal to zero for $\delta V$

= 0 at $T_g = T_{Br-}$ where $\Delta\varepsilon_{lg}(T_{Br-}) = 0$. Then, $(T_{Br-})$ remains unchanged under pressure in the absence of latent heat for $\delta V = 0$. In the case of a first-order transition, $T_g$ is increased and given by Eq. (10), in which $\Delta\varepsilon$ (for $\Pi_2 = 0$) or $\Pi_2$ (for $\Delta\varepsilon = 0$) is the latent heat coefficient when a first-order transition is possible [51].

The underlying first-order transition at $T_{K2}$ without latent heat leads to Phase 3 due to an underlying enthalpy coefficient decrease equal to $\Delta\varepsilon_{lg0}$. When heated above the glass transition $\theta_g$ at zero pressure, the glass Phase 3 transforms into the "ordered" liquid Phase 3 [51,53], which superheats above $T_g$ and $T_m$. It "melts" at the reduced temperature $\theta_{n+} > 0$ given by Eq. (5) and can reappear at $\theta_{n+} < 0$ by homogeneous nucleation of critical superclusters grown around residual superclusters which have not been melted by a high overheating far above $\theta_{n+} > 0$ [48,79]. Equation (20) is used to calculate $\theta_{n+} = \Delta\varepsilon_{lg}$ with $\Delta\varepsilon = 0$ and $(\Pi_1-\Pi_2) = 0$ in agreement with Eq. (5). The nucleation temperature $(\theta_{n+} < 0)$ depends on the overheating rate of liquids above $T_m$ and is expected to disappear when superclusters surviving above $T_m$ are completely melted after an overheating far above $\theta_{n+} > 0$. The number of these residual nuclei must be equal to the number of critical nuclei to be able to observe this transition at $\theta_{n+} < 0$.

The enthalpy coefficient of Phase 3 at $T_m$ given by Eq. (20) for $\theta = 0$ is equal to $\Delta\varepsilon_{lg0} = (\varepsilon_{ls0}-\varepsilon_{gs0})$, which reveals the presence of an underlying first-order transition at $T_{K2}$ in the glass state without latent heat. For water, there is an enthalpy excess $(\varepsilon_{ls0}-\varepsilon_{gs0} = \Delta\varepsilon_{lg0})$ in Phase 3 at $T_m$. This excess is used to predict the value of $T_{K2}$ using Eq. (20), for which $\Delta\varepsilon_{lg} = -\Delta\varepsilon_{lg0}$. Liquids 1 and 2 have different enthalpy coefficients above $T_m$, obeying Eqs. (7,8). There is a transition from Liquid 1 to Liquid 2 at $T_{Br+}$ and $T_{Br-}$, the temperatures where $\varepsilon_{ls}$ becomes equal to $\varepsilon_{gs}$ below and above $T_m$ [41]. This crossover has been observed on Vit 1 in 2007, studying the liquid viscosity above $T_m$ and around $T_{Br+}$ [79]. The crossover at $T_g < T_{Br-} < T_m$ has been recently observed in the fragile liquid $Au_{49}Cu_{26.9}Si_{16.3}Ag_{5.5}Pd_{2.3}$ [80] and cannot be observed in a strong liquid because $T_{Br-} = T_g$. These crossovers could only occur when glass Phase 3 is formed by previous cooling the melt below $T_g$ and further heating above $T_g$ and $T_m$.

## 2.4. The Kauzmann temperature of Phase 3

The entropy $\Delta S(T)$ of Phase 3 is calculated from the specific heat $d(\Delta\varepsilon_{lg})/dT \times \Delta H_m$ and is given in Eq. (22):

$$\Delta S(T) = -2\left(\frac{\varepsilon_{ls0}}{\theta_{0m}^2} - \frac{\varepsilon_{gs0}}{\theta_{0g}^2}\right)\Delta S_m \frac{(T_m-T)}{T_m} + 2(\frac{\varepsilon_{ls0}}{\theta_{0m}^2} - \frac{\varepsilon_{gs0}}{\theta_{0g}^2})\Delta S_m \text{Ln}\left(\frac{T_m}{T}\right). \tag{22}$$

The Phase 3 Kauzmann temperature $T_{K1}$ is the temperature where $\Delta S(T_{K1}) = -\Delta H_m/T_m = -\Delta S_m$. This quantity is not the classical difference between supercooled liquid and crystal entropies because the heat capacity of water is much higher than that of ice at $T_m$ instead of that of Phase 3 which tends to zero at $T_m$.

### *2.6. First-order phase transitions toward the glacial phase and its formation rule*

First-order phase transitions have been observed at normal pressure above $T_g$ in the strong liquid n-butanol [45,81], and the fragile liquids of triphenyl phosphite [43-44] and d-mannitol [46]. Their latent heats decrease the enthalpy coefficients of these phases down to the unique minimum value of $\Delta\varepsilon_{lg}$ ($\theta$) which occurs at the reduced temperature $\theta_{0m}$ and is equal to ($\Delta\varepsilon_{lg}$ ($\theta_{0m}$)) + $\Delta\varepsilon_{lg0}$)×$\Delta H_m$ because the underlying first-order transition acting in all glasses is present and already reduces the initial Phase 3 enthalpy coefficient from zero to (-$\Delta\varepsilon_{lg0}$) [38]. The water first-order transition at normal pressure is viewed in this paper as a fragile-to-strong liquid transformation leading to the strong glacial Phase 3. The first-order transition under pressure of HDA-to-LDA leads in water to the same glacial phase having its enthalpy coefficient equal to the minimum value $\Delta\varepsilon_{lg}(\theta_{0m})$ [51,53]. The HDA phase under normal pressure has its own glass transition lower than that of LDA [56]. The underlying first-order transition without latent heat is controlled by the enthalpy of Phase 3 at $T_{K2}$. The first-order transition temperature leading to the glacial Phase 3 is controlled by its entropy, with its lowest enthalpy being equal to $\Delta\varepsilon_{lg}(T_{0m})\times\Delta H_m$.

### *2.6. Transformation temperature ($T_{sg}$) under pressure of Phase 3 in a glacial phase having a much lower enthalpy*

The pressure varies the enthalpy of Phase 3 as shown in Eq. (20), and induces a new glass phase with a value of $\Delta\varepsilon_{lg}$ equal to zero at the starting temperature $\theta_{sg}$ of the transformation depending on ($\Delta\Pi = \Pi_1-\Pi_2$) and on the value of the enthalpy coefficient. The enthalpy difference after the first-order transition at $\theta_{sg}$ is limited for all pressures by the enthalpy coefficient minimum ($\Delta\varepsilon_{lg}(\theta_{0m})$) obtained for $\Delta\varepsilon = 0$ and $\Delta\Pi = 0$ at the reduced temperature $\theta_{0m}$. The first-order transition at $T_{sg}$ gives rise to a glass phase having a glass transition temperature $T_g$ given by Eq. (10) with $\Delta\varepsilon = |\Delta\varepsilon_{lg}(\theta_{0m})|$ [51]. The enthalpy coefficient $\Delta\varepsilon_{lg}(\theta_{sg})$, being equal to zero at the

departure of the transition under pressure, is submitted to a change equal to E from its initial value without pressure at the new temperature T = $T_g$.

The reduced transformation temperature ($\theta_{sg}$) given in Eq. (23), leading to $\Delta\varepsilon_{lg} = 0$, depends on the value of ($\Delta\Pi_1 = \Pi_1 - \Pi_2$) and on the enthalpy coefficient at the departure of the transition given by Eq. (20) in which $\Delta\varepsilon_{lg0}$ is replaced by ($\Delta\varepsilon_{lg0}$ + E) [53]:

$$\theta sg = -[(\Delta\varepsilon_{lg0} + E + \Delta\Pi_1 + \Delta\varepsilon)]^{\frac{1}{2}} \left(\frac{\varepsilon_{ls0}}{\theta_{0m}^2} - \frac{\varepsilon_{gs0}}{\theta_{0g}^2}\right)^{-\frac{1}{2}}. \tag{23}$$

### 3. Application to water at normal pressure

In this model, the glass transition occurs at $T_g$ = 137.12 K ($\theta_g$ = -0.498) and the first-order transition temperature $T_{LL}$ is fixed at 228.54 K under normal pressure, ($\theta_{LL} = \theta_{Br-}$ = -0.1633) being the reduced temperature where the fragile liquids are transformed into strong liquids under pressure as shown in Chap 4.2. The heat capacity $C_p$ = 104.36 J/K/mol at $\theta_{LL}$ is determined by the initial assumption that the reduced temperature $\theta_{Br-}$ does not depend on pressure because $\Delta\varepsilon_{lg}$ is initially equal to zero in the fragile state before the first-order transition, and $\delta$V P/$\Delta$H$_m$ = 0 at $\theta_{LL}$. The strong supercooled liquid is viewed as a glacial phase induced at $\theta_{LL}$.

*3.1 Enthalpy coefficients of fragile, glacial and strong Phases 3*

The reduced temperatures ($\theta_{0g}$) and ($\theta_{0m}$) are equal to -1 and -2/3, respectively, because the temperatures $T_{0g}$ and $T_{0m}$ are assumed to be equal to 0 K below $T_g$ for strong Liquid 2 and $T_m$/3 for strong Liquid 1, as already shown for pure liquid elements [49,50]. With $\theta_g$ = (-0.498), ($\varepsilon_{gs0}$) is equal to 0.67288 and ($\varepsilon_{ls0}$) to 1.1448 using Eqs. (13,14). The enthalpy coefficients of strong Liquid 1, strong Liquid 2 and Phase 3 obey Eqs. (24-26) after the application of Eqs. (7,8,17) with $\Pi_1 = \Pi_2 = \Delta\varepsilon = 0$:

$$\varepsilon_{ls} = 1.14482 \times (1 - 2.25 \times \theta^2), \tag{24}$$
$$\varepsilon_{gs} = 0.67288 \times (1 - \theta^2), \tag{25}$$
$$\Delta\varepsilon_{lg} = \varepsilon_{ls} - \varepsilon_{gs} = 0.47194 - 1.903 \times \theta^2. \tag{26}$$

Equations (15-18,20) are applied to fragile Liquids 1 and 2 at the reduced temperature $\theta_{Br-} = \theta_{LL}$ = (-0.1633) where $\Delta\varepsilon_{lg}$ = 0 with $\theta_g$ = -0.20305 and the critical free energy barrier in Eq. (21) is infinite. The coefficient (a) in Eq. (15) is equal 0.97 leading to $\Delta C_p$ ($\theta_{LL}$) =29 J/K/mol calculated with a liquid heat capacity of 75.36 J/K/mol at $T_m$, $\Delta H_m$ = 6000 J/mol, and $C_p \cong$ 104.36 J/K/mol. The enthalpy coefficients of fragile Liquid 1, fragile Liquid 2 and Phase 3 are given in Eqs. (27,28,29):

$$\varepsilon_{ls} = 1.80304 \times (1 - 6.336 \times \theta^2), \tag{27}$$

$$\varepsilon_{gs} = 1.69543 \times (1 - 4.357 \times \theta^2), \tag{28}$$

$$\Delta\varepsilon_{lg} = \varepsilon_{ls} - \varepsilon_{gs} = 0.10762 - 4.0366 \times \theta^2. \tag{29}$$

The coefficient a = 0.97 leads to a minimum enthalpy coefficient $\Delta\varepsilon_{lg}(\theta_{0m})$ = -0.52945 in Eq. (29) and along FG in Figure 1 for $\theta_{0m}^2$ = 0.15783. The temperature $T_{0m}$ is equal to 164.6 K in good agreement with a VFT temperature of water equal to 168.9 K [82]. The fragile Phase 3 coefficient $\Delta\varepsilon_{lg0}$ is equal to 0.10762 in Eq. (29) and along AB in Figure 1. Then, the latent heat coefficient for the glacial phase formation must be (-0.52945+0.10762 = -0.421) along CD in Figure 1 while the strong Phase 3 coefficient $\Delta\varepsilon_{lg}$ becomes equal to 0.421 along DE [38]. All these enthalpy coefficient changes lead to a strong glacial phase having a minimum enthalpy coefficient $\Delta\varepsilon_{lg}$ = -0.37382 along HI when $\varepsilon_{ls}$ = 0 at $\theta$ = -2/3.

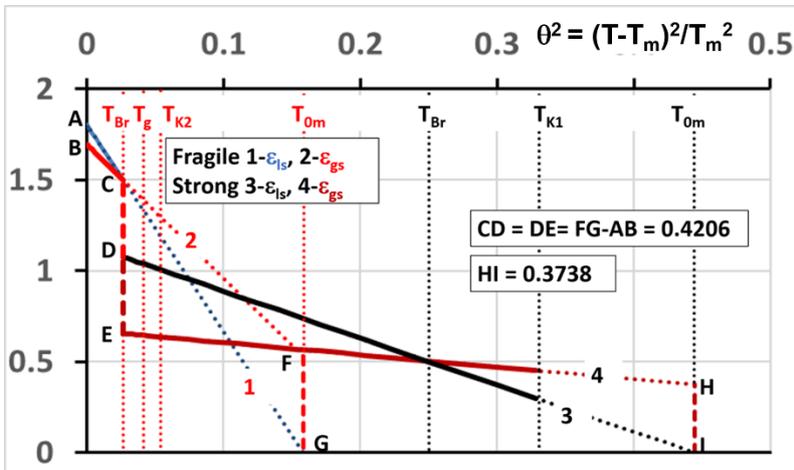

**Figure 1**: Strong and glacial phase formation from fragile liquids 1 and 2 at $\theta_{LL}$ = -0.1633. Enthalpy coefficients $\varepsilon_{ls}$ and $\varepsilon_{gs}$ of fragile and strong liquids 1 and 2 versus $\theta^2 = (T-T_m)^2/T_m^2$. The enthalpy coefficient minimum $\varepsilon_{ls}$ = 0 of the strong Liquid 1 at the point I

is equal to that of the fragile Liquid 1 at the point G. Those of fragile and glacial Phases 3 are $\Delta\varepsilon_{lg}$ = -0.529 along FG and $\Delta\varepsilon_{lg}$ = -0.37382 along HI respectively.

### 3.2. *Heat capacity and entropy of supercooled water*

The heat capacity of supercooled Phase 3 equal to $d(\Delta\varepsilon_{lg})/dT \times \Delta H_m$ using the derivatives of Eq. (26,29) is represented in Figure 2, added to those of hexagonal and cubic ices from the Kauzmann temperature $T_{K1} = 119.8$ K up to $T_m$ [3,9]. The experimental results are plotted around $T_g$ [3] and above $T_{LL}$ [9,10,13]. The total heat capacity is plotted from 100 K to $T_m$. There is a rough agreement with the specific heat values obtained by deducing the contribution of interfacial water at $T_g$ and around $T_{LL}$ and using water in a confined space of 3.3 nm [9,51,53]. The heat capacity change at $T_{LL}$ is equal to 58.6 J/K/mol in Figure 2. The derivative of the enthalpy coefficient difference between Eqs. (27) and (24) and multiplying by $\Delta H_m = 6000$ J/mole leads to a comparable change of 63.5 J/K/mol characterizing a second-order phase transition by heating.

The heat capacity of water is known down to 227 K. Some experimental points are represented in Figure 3 [9-13]. The line joining the experimental points is a polynomial of 5$^{th}$ order leading to $\Delta C_p = 0.25$ JK$^{-1}$mol$^{-1}$ instead of 0 at T = 273.14 K and 29.1 instead of 29 J/K/mole at T = 228.54 K. The $\Delta C_p$ of Phase 3 represented in Figures 2 and 3 is given by the derivative $d(\Delta\varepsilon_{lg})/dT$ of Eq. (29) multiplied by the melting enthalpy $\Delta H_m$ and is equal to zero at $T_m$. The gray area in Figure 2 is the water enthalpy that is not involved in the enthalpy of Phase 3. The heat capacity of Phase 3 and water are not equal because Phase 3, is not yet "ordered" when water is cooled for the first time after vapor condensation.

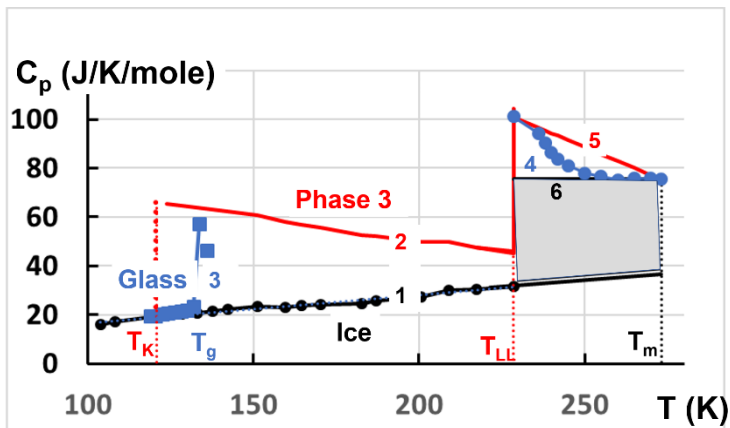

**Figure 2.** Phase 3, hexagonal ice, and experimental [3,9,13] heat capacities of supercooled water. 1- Ice; 2- Phase 3; 3- Experimental points [3]; 4- Experimental points [9,13]; 5- "Ordered" Phase 3; 6- Excess of water heat capacity above $T_{LL}$. Gray area: enthalpy excess of water as compared to that of "ordered" Phase 3 between 5 and 6.

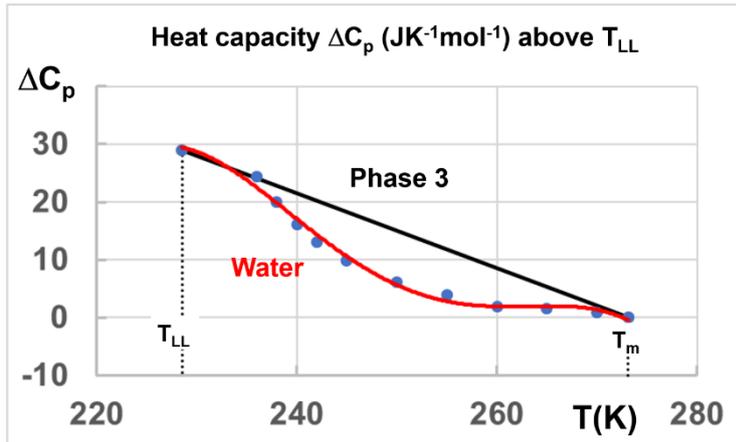

**Figure 3**. Heat capacity of "disordered" Phase 3 [9,13]. "Ordered" Phase 3 is formed along Phase 3 line and is not yet formed on the line of experimental points.

The enthalpy coefficient $\Delta\varepsilon_{lg}$ being 0.10762 at $\theta = 0$ in Eq. (29), its value at $T_{K2} = 122.4$ K, the temperature of the underlying first-order transition temperature without latent heat, is (-0.10762) as shown in Figure 4.

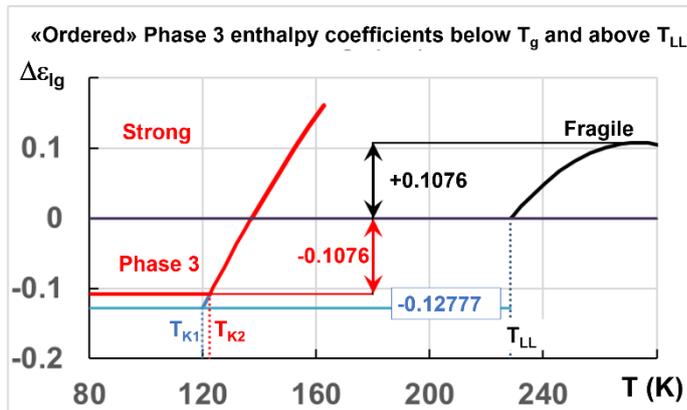

**Figure 4**: Enthalpy coefficients of "ordered" liquid Phase 3. There is a difference of 0.9924 between $\varepsilon_{gs}$ and $\varepsilon_{ls}$ values of fragile and strong phases 3 leading to $\Delta\varepsilon_{lg} = 0$. $\Delta\varepsilon_{lg} = -0.12777$, and $\Delta\varepsilon_{lg} = -0.10762$, the enthalpy coefficients at the Kauzmann temperature $T_{K1}$ and at $T_{K2}$ respectively.

The fragile Phase 3 entropy between $T_m$ and $T_{LL}$ is -2.658 J/K/mole using Eqs. (22,29) when it is "ordered" while the 'experimental' entropy variation between $T_{LL}$ and $T_m$ along Line 1 in Figure 5 is evaluated at equilibrium as equal to -1.5949 J/K/mole from specific heat measurements of water when Phase 3 is not formed. The calculated entropy of the strong liquid along Line 3 at $T_{LL}$ is (-1.2532) J/K/mole using Eq. (22). There is a weak difference of 0.3417 J/K/mole between the prediction and the entropy extrapolated from heat capacity measurements [13]. It can be attributed to an excess of about 2 % in the heat capacity evaluation. The Phase 3 entropy is fixed as equal to the calculated value (-1.2532) J/K/mole at $T_{LL}$ in Figure 5, and equal to $\Delta S_m$ = 21.967 J/K/mole from its Kauzmann temperature $T_{K1}$ = 119.8 K to $T_m$ = 273.14 K. There are entropy and enthalpy differences between those of "ordered" Phase 3 equal to (-1.2532+2.658) = 1.405 J/K/mole and 1.405×$T_{LL}$ = 321 J/mole respectively. A weak pressure reduces this difference and leads to a small "critical" pressure as already proposed [21].

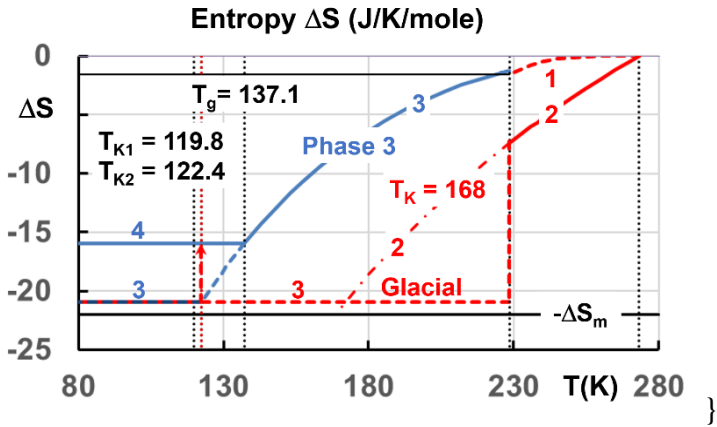

**Figure 5:** Entropy ΔS of various phases. 1- "Experimental" contribution of "disordered" Phase 3 during the first cooling down to $T_{LL}$ = 228.54 K; 2- Water entropy given by Eq. (31) without including contribution N°1; its extrapolated Kauzmann temperature around 168 K; 3- Water transformation at $T_{LL}$ during the first cooling giving rise to the glacial phase LDA that remains liquid with its minimum entropy $\Delta S$ = -20.967 J/K/mole; 4- A first-order transition without latent heat at $T_{K2}$ = 122.4 K inducing "ordering" and vitreous solid by reinforcing enthalpy and density from Line 3 to Line 4; the glass phase, its $T_g$ = 137.1 K and the transition without latent heat of Phase 3 at $T_{LL}$ during heating; the Kauzmann temperature $T_{K1}$ = 119.8 K of Phase 3 with its entropy $\Delta S_m$ = -21.967 J/K/mole equal to that of ice.

The first-order transition at $T_{LL}$ from Line 2 to Line 3 leads to the minimum entropy phase as expected from the glacial phase formation rule and equal to $\Delta S$ = -20.967 J/K/mole as shown in Figure 5. The entropy variation at $T_{LL}$ is 0.529×$\Delta H_m$/$T_{LL}$ = 13.888 J/K/mole with a latent heat equal to FG in Figure 1. The water entropy at $T_{LL}$ along Line 2 is expected to be equal to (-20.967+13.888 = -7.079) J/K/mole in agreement with the experimental-entropy evaluation in Chap. 3.3. Applying Eq. (10) to the fragile liquid 2 with $\Delta \varepsilon$ = 0.529 and $\varepsilon_{gs0}$ = 1.69543 in Eq (28) predicts a glass transition temperature higher than $T_m$. The glacial phase along Line 3 having an entropy still lower remains a liquid below $T_{LL}$ due to the absence of homogeneous nucleation temperature between $T_{LL}$ and $T_m$. A glass phase is expected to be formed at $T_2 = 128$ K in this strong liquid along Line 3 during this first cooling as predicted by Eq. (10) using its enthalpy coefficient $\varepsilon_{gs0} = 0.67288 – 0.10762 = 0.56626$ and $\Delta\varepsilon = 0$ instead of that of fragile liquid. There is no enthalpy change at the homogeneous nucleation temperature $T_2 = 128$ K associated with a glass formation. Enthalpy, entropy and specific heat of glacial phase and of Phase 3 are equal, remain unchanged during this first cooling and are only increasing during subsequent heating at 122.4 K up to Line 4 through a first-order transition without latent heat. Further cooling from liquid Phase 3 gives rise to glass Phase 3 along Line 4 with its underlying first-order transition without latent heat at $T_{K2} = 122.4$ K and its glass transition at 137.1 K as already shown for many other glasses [38]. The glass phase below $T_g = 137.1$ K along Line 4 in Figure 5 cannot be attained by the first cooling and is induced by a second cooling because liquid Phase 3 is now "ordered". This point is emphasized in Chap. 7 entitled 'Origin of "order" in Phase 3'.

### 3.3 The water transformation by first cooling below $T_{LL}$ replaced by Phase 3 formation at $T_{LL}$ = 228.54 K during further cooling.

The heat capacity of water represented by the difference between Line 6 and Line 1 in Figures 2 and 3 is obtained in Eq. (30) by substracting the ice heat capacity from $4.18 \times 18 = 75.24$ J/K/mole:

$$\Delta C_p = -66.93 + 0.1051 \times T \pm 1 (\frac{\frac{J}{K}}{mole}), \tag{30}$$

The associated entropy $\Delta S$ is given by Eq. (31):

$$\Delta S = 0.1051(T_m - T) - 66.93 \ln\left(\frac{T_m}{T}\right) \pm 0.18 \left(\frac{\frac{J}{K}}{mole}\right), \tag{31}$$

leading to $\Delta S = -7.245 \pm 0.18$ J/K/mole for the entropy of Line 2 in Figure 5 at $T_{LL} = 228.54$ K in good agreement with $\Delta S = 7.079$ J/K/mole calculated in Chap. 3.2. The entropy of ice would be attained at a Kauzmann temperature of 168 K far above $T_g = 137.1$ K in the absence of first-order transition at $T_{LL}$. Water undergoes a transition with latent heat at $T_{LL}$ during the first cooling while Phase 3 gives rise to a second-order phase transition at $T_{LL}$ during heating because the negative latent heat of the first-order transition is cancelled by the positive latent heat due to Phase 3 as shown in Part 3.4.

### 3.4. Characteristic temperatures of Phase 3 enthalpy

The second-order transition under Laplace pressure in confined space, also occurs at $T_{LL}$ = 228.54 K, $\theta_{LL}$ = -0.1633, with $T_m$ = 273.14 K [9]. The two reduced temperatures $\theta_g$ and $\theta_{LL}$ where $\Delta\varepsilon_{lg}$ in Eq. (20) is equal to zero with $\Delta\varepsilon = 0$, and $\Pi_1 = \Pi_2 = 0$ cannot depend on pressure because there is no initial volume difference there. In the fragile state of liquids (Phase 3 being "ordered"), these two reduced temperatures $\theta_{LL} = \theta_{Br-} = -0.1633$ and $\theta_{Br+} = +0.1633$ are symmetrical regarding $T_m$ in Figure 6. For $\Delta\varepsilon_{lg} = 0$ above $T_m$, there is a compressibility minimum of bulk water at $\cong 45°C$ where Liquid 2 and Liquid 1 have the same enthalpy coefficient [83,84].

The first-order transition of fragile-to-strong liquid at low pressure only depends on $T_m$ with $\theta_{LL}$ = (-0.1633). The coefficients ($\varepsilon_{ls}$) and ($\varepsilon_{gs}$) of fragile liquids are equal to 1.49841 at $\theta = \theta_{LL}$, while $\varepsilon_{ls}$ and $\Delta\varepsilon_{lg}$ of strong Liquid 1 and Phase 3 are equal to 1.07613 and 0.4212, respectively. In Figure 6, the LLPT at $T_{LL}$ = 228.54 K and room pressure is not accompanied (during heating) by a latent heat. Strong liquid Phase 3 is transformed into "ordered" fragile liquid Phase 3 above $T_{LL}$ which is superheated above $T_m$ and finally "melted" at $\theta_{n+}$ = 0.08108 using Eq. (5). As already shown [20-23,51], a first-order transition at $T_{LL}$ occurs during cooling, producing the transformation of high-density Liquid to low density Liquid in the no man's land.

The enthalpy coefficients in Figure 6 of Liquid 1 on line 1, Liquid 2 on line 2, Phase 3 on line 3, glass phase on line 4 and glacial phase on line 5 are plotted as a function of $\theta$ from -0.7 to zero using Eqs. (24-29). The characteristic temperatures in Kelvin are $T_{0m}$ = 91.05, $T_{K1}$ = 119.8, $T_{K2}$ = 122.40, $T_g$ = 137.12, $T_{LL}$ = 228.54 = $T_{Br-}$, $T_{n+}$ = 250.99, $T_m$ = 273.14, $T_{n+}$ = 295.29, and $T_{Br+}$ = 317.74. Here, the two values of $T_{n+}$ are considered as symmetrical from $T_m$, following previous determinations in other materials [64,65]. There is an enthalpy coefficient $\Delta\varepsilon_{lg}$ equal to zero corresponding to the glass phase along Line 4 with $T_g$ = 137.12 K. The enthalpy coefficients of strong Liquids 1 and 2 are equal to 0.506 at $T_g$ and those of fragile liquids 1 and 2 to 1.49841 at $T_{LL}$.

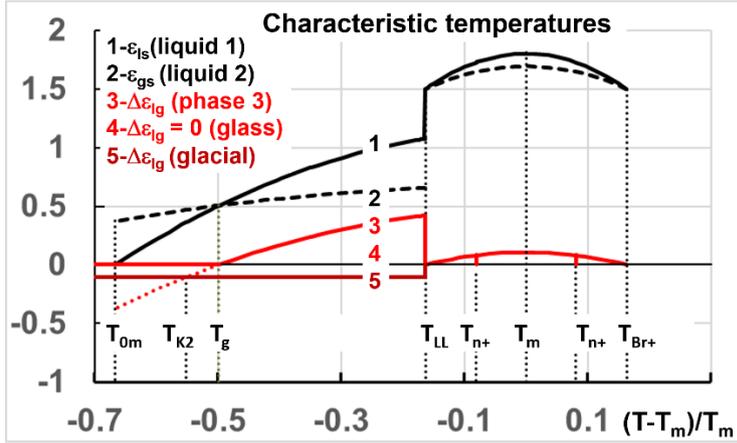

**Figure 6**: Enthalpy coefficient variation versus θ = (T-$T_m$)/$T_m$ and characteristic temperatures: $T_{0m}$, $T_{K2}$, $T_g$, $T_{LL}$, $T_{n+}$, $T_m$, $T_{n+}$, $T_{Br+}$ of Liquids 1 and 2 and Phase 3. $\Delta\varepsilon_{lg}$ = (-0.10762) the lowest enthalpy coefficient of glacial phase along line 5.

The homogeneous nucleation temperatures $T_{n+}$ of fragile Phase 3 given by Eq (5) have not been observed, up to now, due to crystallization of Phase 3. The second temperature $T_{n+}$ = 250.99 K could lead to a transition toward the thermodynamic equilibrium without "ordered" Phase 3 and/(or) to a nucleation of "ordered" Phase 3.

The minimum enthalpy in Eq. (24) is equal to (-0.37382×$\Delta H_m$) for $\varepsilon_{ls}$ = 0 at θ = $\theta_{0m}$ = -2/3 without imposing any entropy constraint related to the Kauzmann paradox [85], as shown in Figures 1. The available enthalpy below $T_g$ is equal to about 1640 J/mole using the crystallization enthalpy near this temperature [3]. This minimum enthalpy cannot be attained with enthalpy considerations because the calculated crystal enthalpy difference of 1402 J/mol and the experimental value are much smaller than 0.37382×6000 = 2243 J/mol. The available enthalpy between $T_K$ and $T_g$ is much smaller [3,51,53].

### 3.5. Application of the formation rule of glacial phases to the HDA-LDA transformation at normal pressure.

A glass transition temperature of HDA has been observed around 116 K under normal atmospheric pressure [56]. An isothermal and sharp enthalpy relaxation has been previously obtained at the same temperature [86] and attributed to the formation of an ultrastable glass phase [53] having a lower enthalpy than that of the glass phase. High-density amorphous ice is still seen as simply a "derailed" state along the Ice I to Ice IV pathway instead of the pure state of HDG [57,58]. The transformation of Ice Ih at 3 kbar when compressed slowly at 100 K, collapses to distorted, tetrahedrally bonded Ice IX while the HDA phase is obtained during a fast compression. These last experiments confirm that HDA and HDG can be a "derailed" state along the ice I to another ice form pathway [59].

The lowest enthalpy of glass state is at once that of the glacial phases of fragile liquids and of HDG as shown in Figures 1 and 6. The pure HDG entropy without "derailed states" is equal to (-20.967+ 0.37382×6000/179.08 = -8.442) J/K/mole below its $T_g$= 116.9 K while its coefficient $\Delta\varepsilon_{lg}$ is given in Eq. (32) using Eqs. (13,14):

$$\Delta\varepsilon_{lg} = \varepsilon_{ls} - \varepsilon_{gs} = 0.62646(1 - 1.6708\times\theta^2) - 0.42211(1 - \theta^2) = 0.20435 - 0.62458\times\theta^2. \tag{32}$$

The minimum enthalpy coefficient of HDG is $\Delta\varepsilon_{lg}$ = -0.16947 for $\theta^2_{0m}$ = 0.59852 in Eq. (32), while that of the low-density glacial and glass phase (LDG) is $\Delta\varepsilon_{lg}$ = (-0.37382) for $\theta^2$ = 4/9 in Eq. (26). Eq. (32) obeys the formation rule of the glacial phase (0.37382-0.16945 = $\Delta\varepsilon_{lg0}$ = 0.62646-0.42211 = 0.20435) [38]. The HDA and LDA enthalpy coefficients are represented in Figure 7 as a function of $\theta^2$ in the absence of a glass transition. The HDA transition temperature is much smaller than 137.12 K and equal to $T_g$ = 116.9 K for $\Delta\varepsilon_{lg}$ = 0 in Eq (32) and in Table 1.

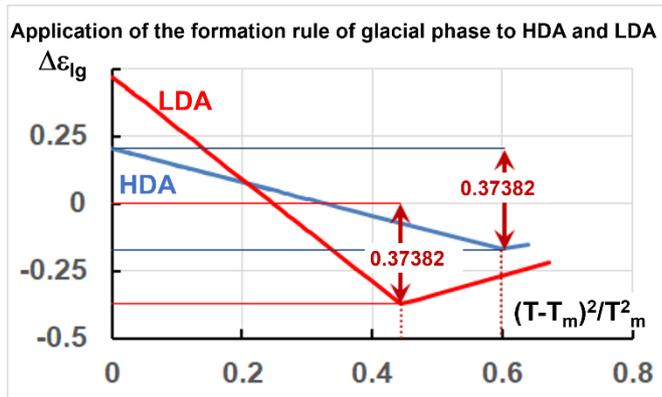

**Figure 7**: Application of the glacial-phase formation rule to the transformation HDA-LDA in the absence of glass transition. LDA and HDA enthalpy coefficients given by Eqs. (26,32) versus $\theta^2$. The minima occur for $\theta = \theta_{0m}$ leading to $\varepsilon_{ls} = 0$ as shown in Figure 1.

The HDA entropy coefficient $\Delta S$ in J/K/mole from $T_{LL}$ to T of "derailed" states HDA and HDG is plotted in Figure 8 along Line 1 with the transformation temperature HDA–LDA occurring at $T_{n-} = T_{01} = 135.1$ K given by Eq. (10) and the latent heat coefficient $\Delta\varepsilon = 0.16947$, $\varepsilon_{gs0} = 0.42211$, $\Pi_2 = 0$, and $\theta_{0g}^2 = 1$ leading to $\varepsilon_{gs} = 0.3143$ in Eq. (32). The first-order transition at $T_{01}$ occurs near the glass transition temperature of the glacial Phase 3 (of LDG).

"Derailed" states are predicted using Eqs. (13,14,20), fixing $T_g$ at 111.8, 112.2, 113.5, 114 and 116.9 K and building the corresponding glass phase giving rise to a glacial phase, varying $\theta_{0m}^2$, ($\varepsilon_{gs0}$) being given by Eq. (13) using $\theta_{0g}^{-2} = 1$ and $\theta_{0m}^{-2}$, $\Delta\varepsilon_{lg0} = \varepsilon_{ls0} - \varepsilon_{gs0}$, with $\varepsilon_{lso}$ determined by Eq. (14). The formation rule of glacial phase being applied, the sum of $\Delta\varepsilon_{lg0}$ and $\Delta\varepsilon$ is equal to 0.37382 for each value of $T_g$ as shown in Table 1. The characteristic parameters are used to predict the entropy of these states below $T_g$ using Eq. (22). "Derailed" HDG phases have an entropy $\Delta S$ from $T_{LL}$ to T given in Table 1 after the transformation of LDA-HDA under pressure which is only 12% or less, smaller than that of pure HDA. HDA has minimum values of $T_g$ and $T_{01}$ equal to 111.8 K and 127.7 K when the LDA entropy is initially equal to that of ice at the beginning of the transformation under pressure. The enthalpy coefficient difference $\delta(\Delta S)$ associated with the formation of orientational disorder [57,58] in HDG (or of melted nanocrystals under pressure and composite phases) is weak. The parameters given in Table 1 for $\delta(\Delta S) = 0$ assumes that LDG has an entropy equal to that of crystals (-21.967) J/K/mole before its transformation under pressure in glass ice with an entropy $\Delta S = (-8.189)$ J/K/mole = (-21.967+12.525+1.2532) of HDG; 12.525, the glacial phase formation entropy = 0.37382×6000/$T_g$ with $T_g$ = 179.08 K and -1.2532 the entropy origin at $T_{LL}$. The LDA entropy depends on the time used to form the glacial phase as already observed for triphenyl phosphite [44]. The processing time increases the nanocrystal fraction (1-f) in LDA before its transformation in HDA. The weak difference of pure HDG with that of vitreous ice equal to 1 J/K/mole and decreasing in each composite as shown in Table 1, is used to evaluate the fraction f of LDG leading to the same entropy variation. The specific heat jump $\Delta C_p(\theta_g) = d(\Delta\varepsilon_{lg})/dT \times \Delta H_m$ is predicted as varying from 15.7 J/K/mole to zero for 111.8 < $T_g$ < 116.9 K in good agreement with the experimental observations [56-58]. The specific heat jump does not exist at $T_g$ = 111.8 K because LDG, having an entropy equal to that of ice before its transformation in HDG, is only characterized by its latent heat of transformation $\Delta\varepsilon \times \Delta H_m$.

The experimental transformation temperatures $T_{01}$ of HDG-LDG are indicated in Figure 8 and correspond to latent heat coefficients $\Delta\varepsilon$ calculated with Eq. (10) in agreement with experimental observations. The entropy behavior confirms the possible existence of orientational disorder of water molecules in HDA or the presence of melted nanocrystals in composite phases [57,58]. The composite character of HDA increases, when the samples of HDA are too slowly prepared and are annealed under a weak residual pressure [56].

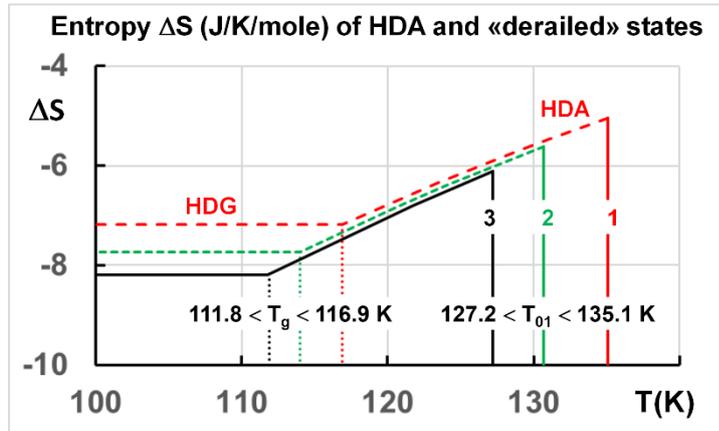

**Figure 8**: Entropy $\Delta S$ in J/K/mole from $T_{LL}$ to T of "derailed" states HDA and HDG. 1- Pure HDG with $\Delta S$ = -7.18 J/K/mole below $T_g$ = 116.9 K, $T_{01}$ = 135.1 K. 2- "Derailed" state with $T_g$ = 114 K, $\Delta S$ =-7.73 J/K/mole and $T_{01}$ = 130.7 K. 3- Entropy $\Delta S$ = -8.18 J/K/mole; minimum of $T_{01}$ = 127.2 K

**Table 1. Glass transition temperature $T_g$, entropy, heat capacity jump at various $T_g$ of HDG containing "derailed" states**

| $T_g$ K | $\varepsilon_{gs0}$ | $\theta_{0m}^{-2}$ | $\Delta\varepsilon_{lg0}$ | $\Delta C_p$ J/K/mol | $\Delta S$ J/K/mol | $\delta(\Delta S)$ J/K/mol | f | f´$\Delta C_p$ J/K/mol | $\Delta\varepsilon$ | $T_{01}$ K | $\Delta\varepsilon+\Delta\varepsilon_{lg0}$ |
|---|---|---|---|---|---|---|---|---|---|---|---|
| 116.9 | 0.42211 | 1.6708 | 0.20435 | 15.7 | -7.18 | 1 | 1 | 15.7 | 0.16947 | 135.6 | 0.37382 |
| 114 | 0.38167 | 1.7037 | 0.21617 | 16.3 | -7.73 | 0.45 | 0.45 | 7.3 | 0.15764 | 130.7 | 0.37381 |
| 113.5 | 0.37504 | 1.7093 | 0.21819 | 16.4 | -7.82 | 0.37 | 0.37 | 6.1 | 0.15563 | 130 | 0.37382 |
| 112.2 | 0.35519 | 1.727 | 0.2243 | 16.7 | -8.1 | 0.08 | 0.08 | 1.3 | 0.14952 | 127.9 | 0.37382 |
| 111.8 | 0.34964 | 1.7322 | 0.22603 | 16.8 | -8.18 | 0 | 0 | 0 | 0.14779 | 127.7 | 0.37382 |

### 3.6. The weak jumps of HDA, LDA and Phase 3 heat capacity at various $T_g$

A sharp glass transition has been observed as shown in Figure 2 with samples of 0.6 g obtained by vapor deposition with a deposition rate of 0.032 g/h [3]. A high heat capacity jump to liquid Phase 3 is expected after vapor deposition when the film is very thick. High jumps are observed for molecular and ionic solutions in which the majority component is water [87]. The surface effects increase the glass transition and reduce the jump height.

The Oguni group used measurements on water in nanoporous media of different pore sizes to eliminate crystallization and to separate the properties of "surface" water from those of "internal" water. The "internal" component shows a heat capacity equal to zero below about 150 K and a progressive increase with temperature due to the distribution of local Laplace pressures up to the second-order phase transition at $T_{LL}$, while the "surface" component has a glass transition near 175 K [9,88].

The vapor deposition is also used to grow multiple nanoscale films of 50 nm thickness alternated with benzene and methanolic films of similar dimensions [6]. The glass transition is now $T_g \cong 179$-180 K because the surface effects could stabilize HDA [51,53]. These results are confirmed on films having a thickness varying from 16 to 110 nm [7]. Using $T_g = 179$ K ($\theta_g =$ -0.34437, $T_m = 273.14$ K), the enthalpy coefficient $\varepsilon_{gs}$ in Eq. (32) is increased by a constant value of $\Delta\varepsilon$ in Eq. (10) leading to $T_g = 179$ K under surface pressure on the thin film. The derivative of the enthalpy coefficient $\Delta H_m \times d(\Delta\varepsilon_{lg})/dT$ of HDA given in Eq. (32) leads to a heat capacity jump at $T_g$ of 9.4 J/K/mole. The measured value is about 6.5 J/K/mole at $T \cong 180$ K using heating rates of $3\times10^4$ K/s [7] at the boundary of the appearance of an exothermic enthalpy associated with crystallization that tends to mask the endothermic enthalpy.

## 4. Application to water under pressure

### 4.1. The line of first-order transitions at $T_{LL} = 0.8367 \times T_m$ for P < 0.017 GPa

At normal pressure, there are entropy and enthalpy differences of 1.405 J/K/mole and 321.1 J/mole ($\Pi = 0.0535$) at $\theta_{LL}$ between fragile and strong Phase 3. A "critical" pressure of 0.0172 GPa = (0.0535×6000/18×0.96×10$^6$ + $P_0$) reduces this difference to zero, assuming a density of 0.96 g/cm$^3$ [21,89] in agreement with another theoretical prediction of 0.018 GPa [21].

A negative pressure pushes the glacial phase enthalpy toward that of crystal. The enthalpy coefficient difference at room pressure is 0.01332. Consequently, a negative pressure of ($\cong$ -0.004 + $P_0$ = -0.0039) GPa crystallizes the glacial phase at 273.44×0.8367 = 228.78 K or creates a "derailed" state or a composite LDA-crystals for -0.0039 < P < 0 GPa.

The $C_p$ increase below $T_m$ proves that the LLPT at $T_{LL}$ is present at normal pressure. The higher limit of the line $T_{LL}(P)$ occurs for P = 0 with the formation of a pure LDA without nanocrystal.

### 4.2. The homogeneous nucleation temperatures of strong liquids under pressure

Pressure moves all the curves represented in Figure 6 along the vertical axis without changing the value of $\theta_{Br-}$. The homogeneous nucleation temperatures $T_1$ and $T_2$ in Figure 9 are calculated as a function of $\Pi_1$ and $\Pi_2$ with Eqs. (10,12), $\Delta\varepsilon = 0$ and $T_{Br-}= 137.12$ K. The homogeneous nucleation temperature $T_1$ of strong Liquid 1 vanishes at $\theta_{LL} = -0.1633$ for $\Pi > 0.85523$ because $T_1$ cannot be larger than $T_m$, while $T_2$ of Liquid 2, is larger than $T_{LL}$ and vanishes for $\Pi_2 = 1.32713$. The supercooled water becomes a strong liquid above $\theta_{LL} = -0.1633$ for $\Pi_2 \approx 0.85523$ ($P \cong \Pi_2 \Delta H_m/V_m = 0.31$ GPa with $V_m = 16.6\times10^{-6}$ m$^3$/mole) up to $\Pi_2 = 1.32713$ ($P \cong 0.55$ GPa). The two strong liquids cannot crystallize above 0.55 GPa in the same ice phase. They give rise to HDG for $0.31 < P < 0.55$ GPa below $\theta_g = -0.34437$ deduced from Eq. (10) with $\Pi_2 = 0.37382$, $\theta_{0g}^2 = 1$, $\varepsilon_{gso} = 0.67288$ and $\Delta\varepsilon = 0$.

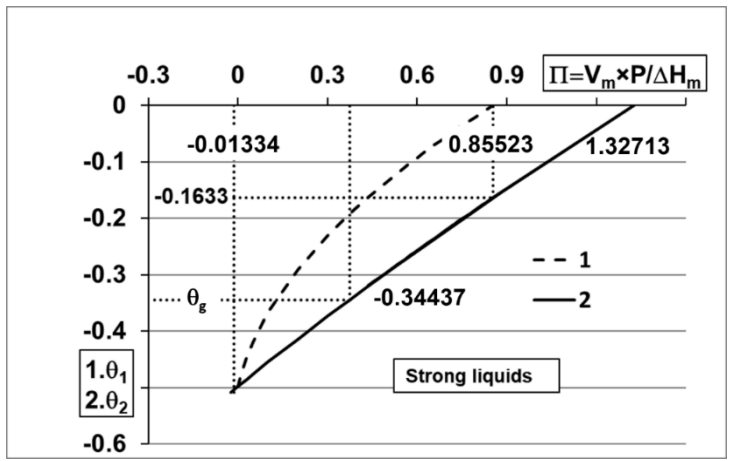

**Figure 9**. Phase diagram of supercooled water and strong liquids under pressure. Curves numbered from 1 to 2: 1– Reduced homogeneous nucleation temperature ($\theta_1$) of strong Liquid 1 vs. enthalpy coefficient $\Pi$ induced by pressure P; 2– Reduced homogeneous nucleation temperature ($\theta_2$) of strong Liquid 2 vs. enthalpy coefficient $\Pi$ induced by pressure P. $\theta_{LL}$ = (-0.1633) for $\theta_1 = 0$ and $\theta_2 = (-0.1633)$.

### 4.3. *The homogeneous nucleation temperatures of fragile liquids under pressure*

The homogeneous nucleation reduced temperatures $\theta_1$ and $\theta_2$ of fragile liquids are calculated as a function of $\Pi_1$ and $\Pi_2$ with Eqs. (10,12) and represented in Figure 10. The highest reduced nucleation temperatures $\theta_1$ and $\theta_2$ are equal to zero ($T_1 = T_2 = T_m$) for $\Pi_1 = 0.19698$ and $\Pi_2 = 0.3046$ respectively while the lowest are always present. These two characteristic pressures equal to about 0.104 and 0.063 GPa have for consequences that Liquids 2 and 1 do not crystallize above θ$_{LL}$ for P > 0.104 GPa under rapid cooling. There is perhaps an opportunity to observe the glass transition at θ$_{LL}$ without crystallization applying pressures with 0.104 < P < 0.55 GPa after an initial cooling below θ$_2$(P). At negative pressures, there is crystallization near θ$_{LL}$ [90].

The temperatures θ$_1$ and θ$_2$ are equal to θ$_{LL}$ = -0.1633 for Π = 0.01168 (P ≅ 0.0035 + P$_0$ ≅ 0.0036 GPa) in Figure 10. A new glass transition temperature appears in the fragile state above the initial θ$_g$ = -0.20305 because the pressure induces an enthalpy coefficient equal to 0.01168, added to ε$_{ls}$ and ε$_{gs}$, leading to Δε$_{lg}$ = 0 at θ$_{LL}$. Second glass transition temperatures are also observed in stable glasses prepared by vapor deposition because the same enthalpy excess coefficient Δε is added during a rapid heating to the coefficients ε$_{ls}$ and ε$_{gs}$ of fragile Liquids 1 and 2, obeying Eq. (10) [91 and references therein].

The reduced temperatures θ$_1$ and θ$_2$ are larger than θ$_{LL}$ = -0.1633 for 0.01168 < Π < 0.19698 without changing the glass transition temperature which occurs at θ$_{LL}$ for Δε$_{lg}$ = 0. The first-order glass transition temperature θ$_{LL}$ (Π) = -0.1633 still exists for 0.01168 < Π < 0.85523 after cooling and reheating, and is higher than the formation temperature of the glacial phase for 0.017 < P < 0.31 GPa and accompanied by a heat capacity jump of 29 J/K/mole.

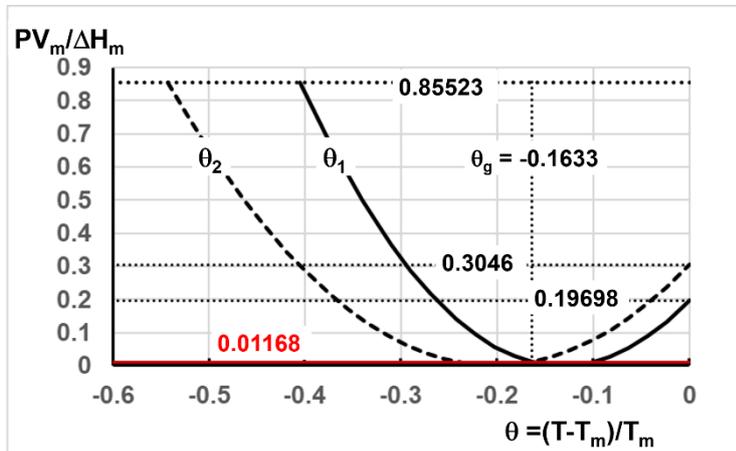

**Figure 10**. Homogeneous nucleation temperatures θ$_1$ and θ$_2$ of fragile Liquids 1 and 2. Line of glass transitions θ$_{LL}$ = θ$_g$ = -0.1633. The upper limit Π = 0.85523 for the first-order transition fragile-to-strong liquids only acting after the first cooling to low temperatures under pressure followed by reheating. Line Π = 0.01168 above which θ$_g$ = -0.1633.

*4.4. The first-order transformations for 0.0535 < Π < 0.85523 (0.017 < P < 0.31 GPa).*

The negative enthalpy coefficient of fragile Phase 3 under pressure is plotted in Figure 11 as a function of θ using Eq. (29) and deducing 0.0535 to obtain $\Delta\varepsilon_{lg} = 0$ under a pressure of 0.017 GPa. There is a jump of $\Delta\varepsilon_{lg}$ from -0.0535 to zero for $\Pi_1 = 0.0535$ at $\theta_{LL} = -0.1633$. The first-order transition reduced temperature decreases under pressure and occurs at each reduced temperature $\theta_{sg}$, where $(\Delta\varepsilon_{lg} + \Delta\Pi_1) = 0$ for each value of $\Delta\Pi_1 = \Pi_1 - \Pi_2 = -\Delta\varepsilon_{lg}$. A line of first-order transitions is expected because there is a dependence of the isothermal compressibility toward a critical point $T_S$ under various pressures [54]. The extrapolated values are $T_S = 175$ K for P = 0.19 GPa and $T_m = 254$ K; $T_S = 192$ K for 0.15 GPa and $T_m = 259.1$ K; $T_S = 212$ K for 0.1 GPa and $T_m = 264.4$ K; $T_S = 224$ K for P = 0.05 GPa and $T_m = 269.1$ K. The corresponding melting temperatures obey the Clausius-Clapeyron law $dP/dT_m = 0.0134$ at low pressure and Eq. (33) at pressure less than 0.5 GPa [92]:

$$T_m(K) = 557.2 - 273\exp\left(\frac{(300+P(MPa))^2}{2270000}\right). \tag{33}$$

These temperatures $T_m$ are used in Figure 11 to determine θ, and $\Delta\varepsilon_{lg}(\theta) = \Delta\Pi_1 = \Pi_1 - \Pi_2$ in agreement with the model. The upper limit of the first-order transition line is 0.31 GPa ($\Pi = 0.85523$) because the enthalpy coefficient of fragile Phase 3 cannot be lower than (-0.42183) in Figure 1 below 154.7 K ($T_m \cong 235.6$ K). The boundary separating the glass transition temperatures of fragile liquids and those of HDA is found at θ = -0.34338 for $\Delta\varepsilon_{lg} = -\Delta\Pi_1 = -0.42183$ and is very close to the reduced glass transition of HDA under pressure at $\theta_g = -0.34437$. The enthalpy coefficient $\Delta\varepsilon_{lg}$ of the glacial phase is given by a constant added to Eq. (26) in order to obtain $\Delta\varepsilon_{lg} = 0.42183$ at the end of the first-order transition. The point of transformation observed at 215 K for P = 0.1 GPa also agrees with these predictions [55].

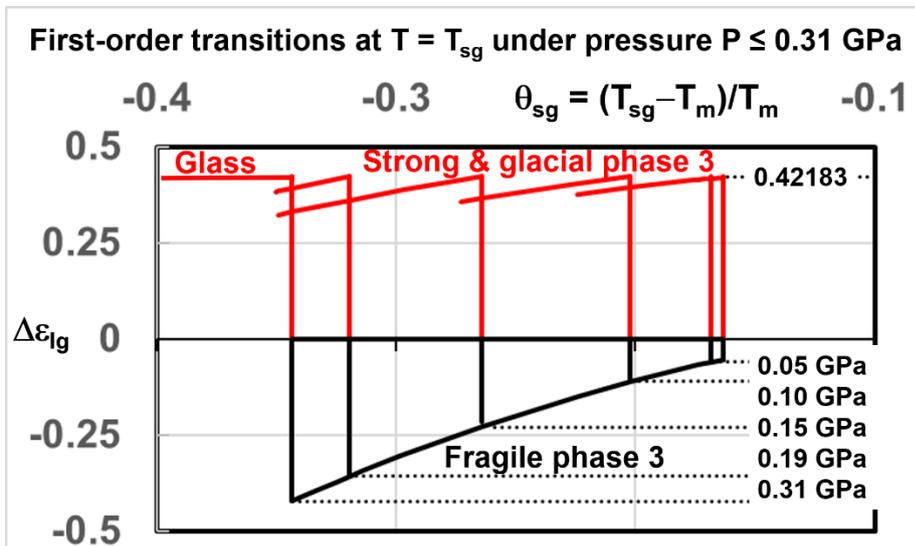

**Figure 11.** First-order transitions from the fragile glass phase at normal pressure toward the glacial phase $\Delta\varepsilon_{lg}$ under pressure $0.018 < P < 0.31$ GPa: Enthalpy coefficients $(\Delta\varepsilon_{lg} + \Delta\Pi_1) = 0$ at the starting temperature $T_{sg}$ of first-order transitions under pressure P with $\Delta\varepsilon_{lg}$ = (Eq. (29) - 0.0535). The end of the first-order transition at $\Delta\varepsilon_{lg} = 0.42183$ belonging to the glacial phase under pressure. The line $(\Delta\varepsilon_{lg} + \Delta\Pi_1) = 0$ of fragile and strong glass phases. $\Delta\varepsilon_{lg}$ varying from (-0.42183) to 0.42183 for $\theta = -0.34338$ and P = 0.31 GPa. $\Delta\varepsilon_{lg}$ varying from zero to 0.42182 for $\theta = -0.1633$ and $P \cong 0.0001$ GPa as described in Figure 6.

### 4.5. The first-order transformations for $0.85523 < \Pi < 1.32713$ ($0.31 < P < 0.55$ GPa)

High pressure applied to ice samples followed by complete decompression induces LDA which has a density and, consequently, an enthalpy close to that of ice. The melting temperature ($T_m$) is strongly decreased under pressure. The compression of LDA transforms it into HDA at a well-defined pressure (P). This sharp transformation is a first-order transition occurring at ($T_{sg}$) defined in Eq. (23) and represented in Figure 12. The volume change ($\delta V_2$) leading to the glacial phase equal to $0.2$-$0.22\times10^{-6}$ cm$^3$/g does not depend on the pressure (P) [16,18].

The glacial phase transition of strong liquid 2 under atmospheric pressure is $\theta_g = -0.34437$ given by Eq. (10) with $\varepsilon_{lg0} = 0.67288$, $\Pi_2 = 0$, $\theta_{0g}^2 = 1$ and $\Delta\varepsilon = 0.37382$ following the formation rule of glacial phase with $\Delta\varepsilon$ = HI in Figure 1. The enthalpy coefficient $\Delta\varepsilon_{lg}$ ($\theta$) of Phase 3, initially equal to 0.24627 along Line 2 at $\theta_g$, is equal to zero at the departure of the first-order transition under pressure. Consequently, $\Delta\varepsilon_{lg0}$ in Eqs. (20,23,26) is replaced by $(\Delta\varepsilon_{lg0} - E = 0.22567)$ with $E = -0.24627$ and gives rise to the enthalpy coefficient of glacial phase on Line 4. The pressure induces an enthalpy coefficient change $\Delta\Pi_1$ transforming Line 4 in Line 3 respecting $\Delta\varepsilon_{lg} + \Delta\Pi_1 = 0$ in Figure 12 at the departure of the first-order transition and finally leading to $\Delta\varepsilon_{lg} = 0.37382$ on Line 1. The enthalpy coefficient variation $\Delta\Pi_2 = 0.37382$, respects the glacial phase formation rule and leads to a constant volume variation. The first-order glass transition occurs at various $T_g = 0.65563 \times T_m(P)$ during reheating under pressure and leads to the liquid enthalpy coefficient for each value of P.

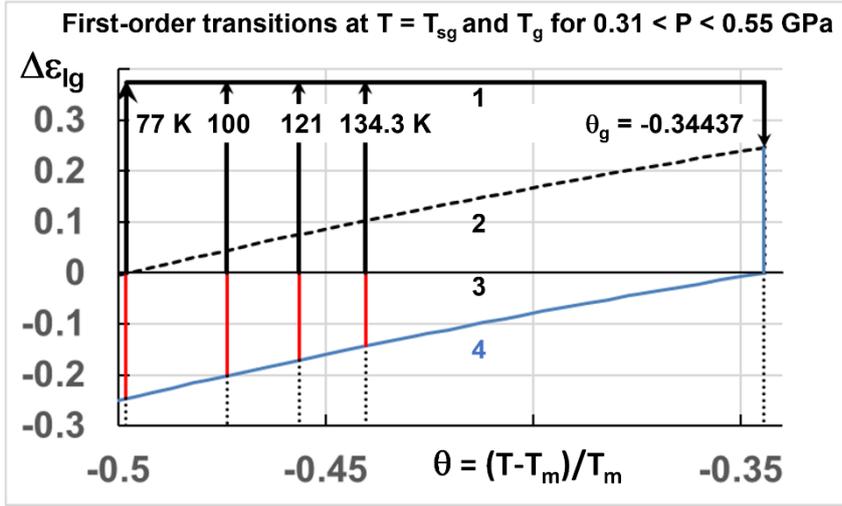

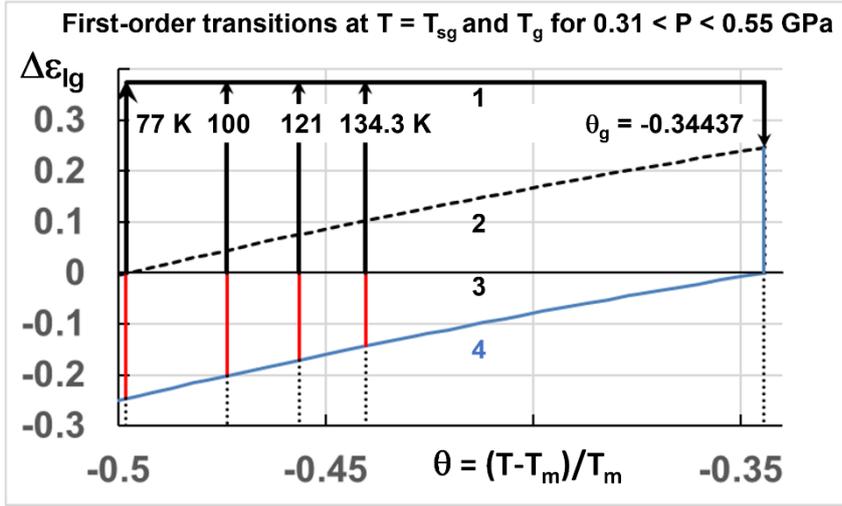

**Figure 12.** Enthalpy coefficients $\Delta\varepsilon_{lg}$ at the starting temperature $T_{sg}$ of first-order transitions under pressure P and at the end of the transition. $\Delta\varepsilon_{lg}$ of the strong Phase 3 given by Eq. (26) along Line 2; $\Delta\varepsilon_{lg}$ given by Eq. (26) in which 0.47194 is replaced by 0.24627 along Line 4; $\Delta\varepsilon_{lg} = 0$, departure of the first-order transition along Line 3. $\Delta\varepsilon_{lg} = 0.37382 = \Delta\Pi_2$ at the end of the transition along Line 1. Temperatures, reduced temperatures and pressures in Table 2. Lines 3 and 4 are separated by $\Delta\Pi_1$, the enthalpy coefficient difference induced by pressure at the departure of the first-order transition.

**Table 2.** First-order glass-to-glass transitions under various pressures [53].

| 1 | P (GPa) | 0.55 | 0.45 | 0.38 | 0.32 |
|---|---|---|---|---|---|
| 2 | $T_{sg}$ (K) | 77 | 100 | 121 | 134.3 |
| 3 | $T_m$ (K) | 153.4 | 190 | 222.6 | 240 |
| 4 | $\Delta\Pi_1$ | 0.24627 | 0.2015 | 0.1706 | 0.14329 |
| 5 | $\theta_{sg}$ | -0.498 | -0.47379 | -0.45633 | -0.44033 |
| 6 | $\Delta\Pi_2$ | 0.37382 | 0.37382 | 0.37382 | 0.37382 |
| 7 | $\theta_g$ | -0.34437 | -0.34437 | -0.34437 | -0.34437 |
| 8 | $T_g$ | 100.6 | 124.7 | 145.9 | 157.4 |

All parameters predicting these phenomena are given in Table 2 and Figure 12 [51,53]. The first-order transitions lead to HDG on Line 1 with $\Delta\varepsilon_{lg} = \Delta\Pi_2 = 0.37382$. The first-order transitions have been observed at $T_{sg}$ = 77 K, 100 K, 121 K, and 133.4 K under pressure P equal to about 0.55, 0.45, 0.38, and 0.32 GPa respectively [16]. The values of $\Delta\Pi_1 = \Pi_1-\Pi_2$ are proportional to the applied pressure. The glass phase disappears for $\Pi_2 > 1.32713$ and Eq. (23) cannot be applied for pressures P > 0.55 GPa.

*4.6. All first-order transition temperatures under pressure including glass transitions*

The line $P(T_{sg})$ represented in Figure 13 is continuous. The glass transition temperatures are $T_g$ = 0.65563×$T_m$(P) for HDA with 0.31< P < 0.55 GPa and 0.8367×$T_m$(P) for fragile glasses with 0.017 < P < 0.31 GPa. The two boundaries $\theta_g$ = -0.34437 and -0.34338 are not perfectly equal because the choice of thermodynamic parameters of liquids is good but imperfect. There is a good agreement between our predictions and all the experimental observations [16,55]. Another prediction of a transformation at P = 0.017 GPa looking like a critical point is confirmed [21].

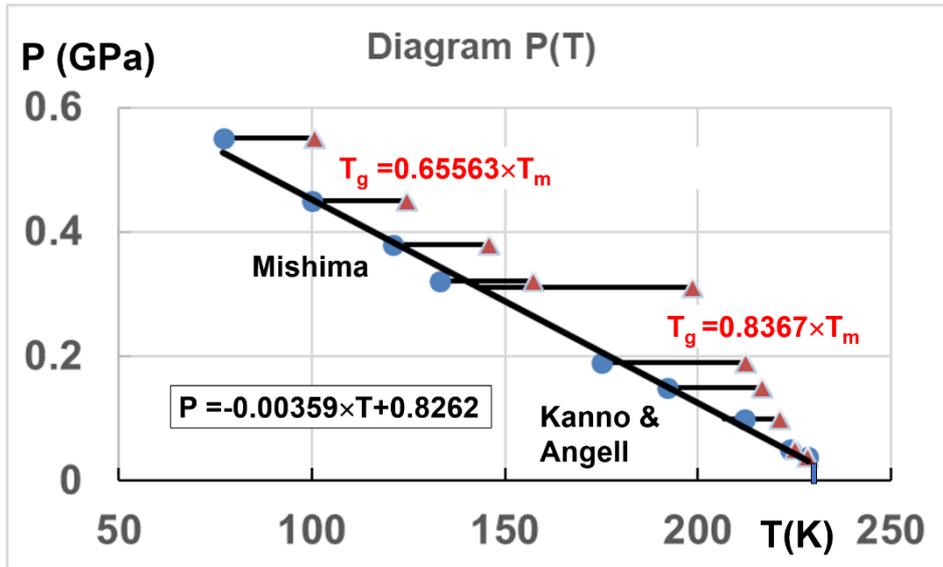

**Figure 13**: First-order transitions and diagrams $P(T_{sg})$ and $P(T_g)$. Circles ●: The first-order transition at $T_{LL}$ exists from 0 to 0.017 GPa. The line $P(\theta_{sg})$ starts from 0.017 GPa at $T_{LL}$ = 227.47 K and has an upper limit of about 0.55 GPa at 77 K. Triangles ▲: The second reduced glass transition temperature of the fragile liquids is $\theta_g$ = -0.1633 ($T_g$ = 0.8367×$T_m$) for 0.0036 < P < 0.31 GPa. Those of strong LDA and HDA under pressure are $\theta_g$ = -0.34437 ($T_g$ = 0.65563×$T_m$) above 0.31 GPa.

*5. The HDA-VHDA transformation at the homogeneous nucleation temperature $T_{n+}$ of superheated Phase 3 for P = 0.95 GPa.*

Two volume changes are observed with the increase of pressure at 125 K. The first one is the LDA-HDA transformation around 0.4-0.5 GPa as discussed in Chap. 4.5. The second volume change, corresponding to HDA-VHDA transformation, occurs under 0.95 GPa [19]. A supplementary volume reduction is obtained after decompression at 77 K and is equal to 0.09 cm$^3$/g. This transition occurs in fact above $T_m$ in the strong liquid when Eq. (5) is respected for $\theta_{n+}$ = 0.3003 ($T_{n+}$ =1.3003 $T_m$) and is accompanied by an endothermic enthalpy equal to 0.3003 ×$\Delta H_m$ =1802 J/mol deduced from Eq. (26) [51,53]. The supplementary and theoretical volume change $\delta V$ =1802/18/P = $10^{-1}$ cm$^3$/g agrees with the experimental results. The melting temperature $T_m$ of ice Ih, under 0.95 GPa, is deduced as equal to 125/1.3003 = 96 K in rough agreement with Mishima's measurements of $T_m$ [17]. This transformation at 125 K above $T_m$ is a LLPT. The glass state is obtained by reducing the pressure. This first-order transition above $T_m$ corresponds to the "melting" of the "ordered" liquid Phase 3. The nucleation of this phase at $\theta_{n+}$ = $\Delta\varepsilon_{lg}$ = -0.3003, expected from Eq. (3) during decompression is absent because the enthalpy excess is not recovered. All residual superclusters including their dense cores are melted because the applied pressure above 1.4 GPa induces an ice melting temperature which is much less than 125 K. The supplementary volume reduction of 0.1 cm$^3$/g after decompression is frozen in the absence of this transition [86].

## 6. Enthalpy and entropy of "ordered" Phase 3 above $T_{LL}$ and $T_m$

The thermodynamic properties of Phase 3 are now predicted above $T_{LL}$. The heat capacity of water which contains "disordered" Phase 3 is known and used to compare with Phase 3 properties in Figure 2 and 3. The entropy of Phase 3 below $T_{LL}$ is represented in Figure 5 under room pressure.

### *6.1. Entropy coefficients around $T_m$.*

The entropy coefficients of water and fragile and "ordered" Phases 3 are represented in Figure 14 as a function of the temperature with Eq. (22) on Lines 1 and 3 and the "experimental" entropy of water on Line 2. The "ordered" Phase 3 entropy on Lines 1 and 3 between the two first-order transitions at $T_{n+}$ are separated by a difference corresponding to a latent heat of coefficient 0.08108 with $T_{n+}$ = 251 and 295.28 K. The water entropy along Line 2 is a mixture of Liquids 1 and 2 entropies without "ordered" Phase 3 because the water at thermodynamic equilibrium results from vapor condensation. The "ordered" Phase 3 occurs at $T_g$. This result raises the question of the predicted existence of fragile Phase 3 above T$_{LL}$ and of the calculated nucleation temperatures at T$_{n+}$ if crystallization is avoided.

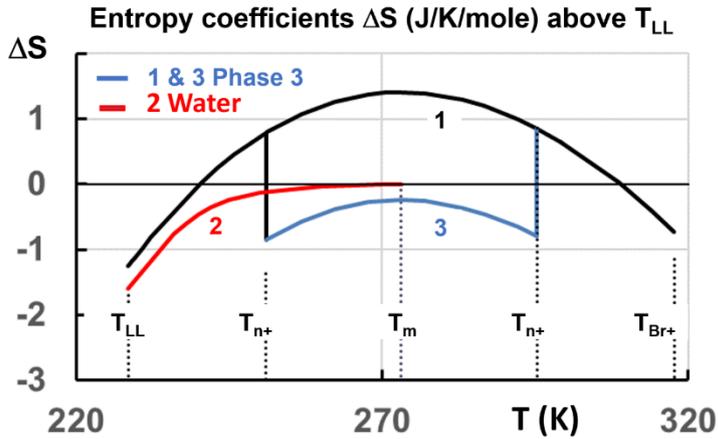

**Figure 14.** Entropy of water and fragile "ordered" Phase 3 between T$_{LL}$ = 227.88 K and T$_{Br+}$ = 317.74 K. Phase 3 entropy along Lines 1 and 3. Water entropy without superclusters along Line 2 (extrapolated from experimental values of $\Delta C_p$ (T) [13]). First-order transitions at T$_{n+}$ = 250 and 295.28 K of fragile Phase 3. The calculated entropy of Phases 3 at T$_{LL}$ is (-1.2532) J/K/mole instead of -1.59 J/K/mole for Line 2.

### 6.2. Enthalpy coefficients between T$_{LL}$ and T$_m$

The enthalpy of fragile and "ordered" Phases 3 is represented in Figure 15 by Lines 1 and 3 with coefficients of 0.10762 and 0.02654 at T$_m$. The two lines 1 and 3 are separated by a difference of enthalpy coefficients equal to 0.08108. The boundaries are the two temperatures T$_{n+}$. "Ordered" supercooled Phase 3 is underlying from T$_g$ up to T$_{n+}$. The lowest enthalpy appears at T$_{n+}$ = 250 K and increases with the temperature up to (0.02654 = 0.10762-0.08108) at T$_m$ as shown along Line 3. The "experimental" enthalpy coefficient variation of water from zero at T$_{LL}$ to (0.06317) at T$_m$ along Line 2 is calculated from the experimental heat capacity in Figure 3 [13]. The water enthalpy appears to be a mixture of Liquid 1 and Liquid 2 enthalpies without "ordered" Phase 3. The two Phases 3 are separated by supercooling and superheating at temperatures T$_{n+}$ which are expected to exist in supercooled and superheated water having escaped from crystallization. These two temperatures have already been observed in other glass-forming melts [38,60,64,65]. These homogeneous nucleation temperatures of fragile liquids above T$_{LL}$ have not been observed in water up to now and could exist after supercooling Phase 3 down to T$_g$ and reheating above T$_{LL}$ without crystallization if it is possible. Nevertheless, the homogeneous nucleation temperature T$_{n+}$ of strong liquids for P = 0.95 GPa has been already observed [19]. These temperatures $T_{n+}$ lead to two first-order transitions at these temperatures, as indicated with arrows in Figure 14.

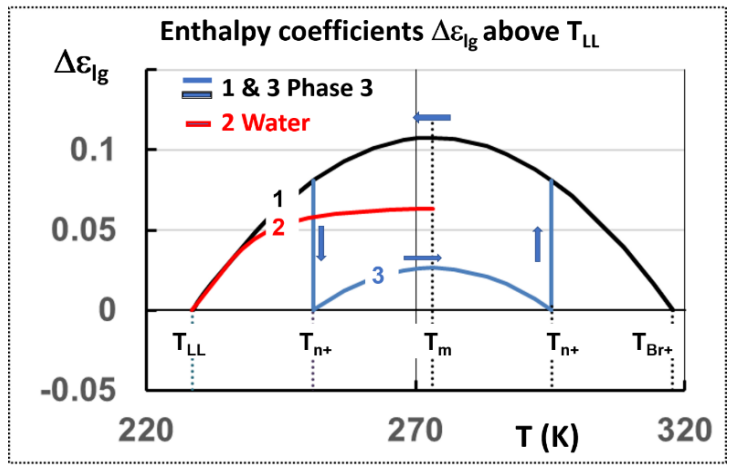

**Figure 15.** Enthalpy coefficient variations of water and "ordered" Phase 3 between $T_{LL}$ = 227.88 K and $T_{Br+}$ = 317 K. Phase 3 enthalpy along separated Lines 1 and 3. Water enthalpy without "ordered" Phase 3 along line 2. Transitions at $T_{n+}$ = 250 and 295.28 K.

## 7. Medium-range order in liquid Phase 3

Water is transformed into a glacial phase by a first-order transition at $T_{LL}$ = 228.5 K during the first cooling if crystallization is avoided as shown in Figure 5. This new phase having the lowest entropy and enthalpy remains a frozen liquid below $T_{LL}$ down to the lowest temperatures. The glass transition is not observable during cooling because its entropy, being minimum, cannot decrease. The glacial liquid phase and Phase 3 are transformed at $T_{K2}$ = 122.4 K in glass during heating, giving rise to the vitreous solid by an increase of density and entropy, accompanying the first-order transition without latent heat. Thermal fluctuations weaken the bonds and the degree of connectivity between atoms with temperature. The glass transition at 137.1 K looks like a percolation-type phase transition during heating in the system of broken bonds called configurons with formation of dynamic fractal structures above the percolation threshold and gives rise to liquid phase 3 with medium-range order [28].

The liquid Phase 3 is ordered above the temperature $T_g$, increases its enthalpy and its density with temperature, and superheats up to $T_{n+}$ = 295 K if crystallization is avoided. A second cooling from temperatures lower than $T_{n+}$ does not modify the medium-range order in liquid Phase 3 at temperatures lower than $T_{LL}$. Liquid Phase 3 is transformed into a vitreous solid at $T_g$ = 137.1 K during the second cooling. The transition at $T_{K2}$ = 122.4 K is underlying below $T_g$ during the second cooling because the strong liquid is already "ordered" thanks to the first cooling. The model of twinkling fractal theory is also compatible with the formation of solid-like clusters and the development of percolating solid fractal structures during further cooling through $T_g$ as experimentally proved via AFM-visualization [29,31].

All atoms in ordered liquid phase 3 are organized in blocks containing many elementary superclusters. These blocks percolate and then grow at $T_2 = T_g$ during the second cooling, give rise to the glass phase, enthalpy freezing and heat capacity jump [28-30,72]. The atom number n in elementary blocks is about 102 at $T_g$ as given in Eq. (19). New blocks are built after reheating characterized by $\Delta\varepsilon_{lg} = 0.08108$ at $T_{n+}$. The ratio $\varepsilon_{ls}(T_g)/\Delta\varepsilon_{lg}(T_{n+})$ is equal to $0.506/0.0818 = 6.24$ where $\Delta\varepsilon_{lg}$ and $\varepsilon_{ls}$ are assumed to be proportional to the Laplace pressure for high values of R and to the reverse of the supercluster radius R. Using this rough approximation, the atom number of superheated blocks could be $(6.24)^3$ times larger than 102 and equal to 24783 before being melted at $T_{n+} = 295$ K. This number is certainly too high. Nevertheless, superheated blocks could contain several thousand atoms before being melted at $T_{n+} = 295.3$ K.

After their melting, these critical blocks could revive during cooling at $T_{n+} = 251$ K because their dense core is only melted after a very high overheating far above $T_{n+} = 295$ K. These two homogeneous nucleation temperatures $T_{n+}$ are symmetrical from $T_m$ because the number of superclusters is unchanged after a moderate overheating above $T_{n+} = 295$ K and consequently the formation enthalpy of each supercluster is equal to its melting enthalpy at $T_{n+} = 251$ K.

## 8. Conclusions and perspectives

The classical nucleation equation applied to two liquids (Liquid 1 and Liquid 2) is completed for all glass-forming melts by additional enthalpy fractions $\varepsilon_{ls}$ for Liquid 1, $\varepsilon_{gs}$ for Liquid 2 of the melting enthalpy $\Delta H_m$. These quantities are linear functions of $\theta^2 = (T-T_m)^2/T_m^2$ that govern the solid supercluster formation during first cooling under Laplace pressure. The homogeneous nucleation temperatures $T_1$ and $T_2$ become transformation temperatures of Liquids 1 and 2. Below these temperatures, "ordered" liquids 1 and 2 are characterized by short-range order. The strong liquids 1 and 2 are transformed by heating into a new phase called Phase 3 through a first-order transition without latent heat at the temperature $T_{K2} < T_2$. It is characterized by an enthalpy coefficient equal to $\Delta\varepsilon_{lg} = (\varepsilon_{ls} - \varepsilon_{gs})$, and composed of percolated and interpenetrated superclusters giving rise to a glass below $T_g$ at the homogeneous nucleation temperature $\theta_2$ of Liquid 2 and to a new "ordered" liquid Phase 3 above $T_g$, characterized by medium-range order which superheats up to $T_{n+}$ above the melting temperature $T_m$. Phase 3 has an enthalpy coefficient equal to $\Delta\varepsilon_{lg0}$ at $T_m$. Its entropy has its own Kauzmann temperature.

The heat capacity difference of undercooled water with ice is 38.2 J/K/mole at $T_m$. Consequently, its extrapolated Kauzmann temperature at 168 K is far above its glass transition at 137.1 K. The formation of Phase 3 with its own Kauzmann temperature is needed to explain this experimental fact.

The model of Phase 3, successfully applied to d-mannitol, triphenyl phosphite and n-butanol, still defines a formation rule of glacial phase, explaining the origin of the first-order transition of water from fragile-to-strong liquid at $T_{LL}$ = 228.5 K, only knowing $T_g$ = 137.1 K, the melting heat $\Delta H_m$ and the melting temperature $T_m$. All thermodynamic properties and transitions of Phase 3, even under pressure P, are now predicted in agreement with experiments of Kanno and Angell (1979), Mishima (1994), Mishima and Stanley (1998), Loerting et al (2006), Amann-Winkel et al (2013), Shephard and Salzmann (2016, 2017), Tulk et al (2019). The glacial phase is formed at $T_{LL}$ = 0.8367×$T_m$ for P < 0.017 GPa. The transformation temperature ($T_{LL}$) decreases with P > 0.017 GPa and disappears for P > 0.55 GPa. The lowest-density liquid is, at once, the glacial phase of fragile and high-density liquids. There are two low-density phases at low temperatures: the glacial phase with the lowest density and the glass phase with a higher density and $T_g$ = 137.1 K. The high-density liquid has a glass transition at $T_g$ = 116.9 K under normal pressure. The glass transition of fragile state accompanied by a weak latent heat is predicted for the first time, as occurring at $T_g$ = 0.8367×$T_m$(P) for 0.017 < P < 0.31 GPa.

The glacial phase formed at $T_{LL}$, remains liquid during the first cooling, and gives rise to strong glass Phase 3 by heating through the first-order transition without latent heat at $T_{K2}$ = 122.4 K. Strong liquid Phase 3, with medium-range order, appears during first heating above $T_g$ = 137.1 K, with its own Kauzmann temperature, which is transformed at $T_{LL}$ into a fragile "ordered" Phase 3 with a new enthalpy coefficient $\Delta\varepsilon_{lg}$, superheating above $T_m$ up to $T_{n+}$ = 295 K if crystallization is avoided. The exothermic latent heat at $T_{LL}$ exists during the first cooling and is compensated during the next cooling by the Phase 3 transformation heat. The first-order transition, without latent heat at $T_{K2}$ = 122.4 K, becomes underlying below $T_g$ in further heating and induces an enthalpy excess at $T_m$ which is used to predict $T_{K2}$.

Precursory signs of this new liquid Phase 3 with medium-range order exist in other substances and in supercooled water under high pressure. Its existence is a direct consequence of the presence of two liquids in these materials. Its successful use for supercooled water still implies the presence of two liquids and the need of a complementary enthalpy in the classical nucleation equation to solve supercooled melt problems.

The homogeneous nucleation temperature at $T_2 = T_g$ in glass-forming melts looks like a percolation-type phase transition without specific heat jump during the first cooling from very high temperatures. The glass transition of Phase 3 appears, during first heating, in the system of broken bonds called configurons with formation of dynamic fractal structures above the percolation threshold and medium-range order in liquid Phase 3 up to $T_{n+}$. The glass transition is observed during further cooling from temperature lower than $T_{n+} > T_m$.

The formation of glacial phases by isothermal annealing above $T_g$ induces composite glasses containing nanocrystals. The annealing time must be minimum to obtain a pure glass phase. The HDA phase of supercooled water under normal pressure obtained after annealing near the transition temperature HDA-LDA around 140 K under a weak residual pressure is analyzed as a composite of glass phases below 116.9 K with a composition depending on annealing time. The derailed states are mixed with pure HDA characterized by an entropy being the sum of those of initial and melted ice and of glacial phase.

**Acknowledgments**: The author thanks Pascal Richet for several suggestions during this work and Michael Ojovan for a comment considering that the absence of glass transition in many other glass-forming melts after hyperquenching is due to the first cooling from very high temperatures.